\documentclass[amsmath,amssymb,11pt]{article}
\usepackage{jheppub}
\pdfoutput=1
\usepackage{graphicx,epsfig}
\usepackage{float}
\usepackage{subfig}
\usepackage{subfloat}
\usepackage[utf8]{inputenc}
\usepackage{hyperref}
\usepackage{caption}
\usepackage{color}
\usepackage{subcaption}
\usepackage{tensor}
\usepackage{cleveref}

\allowdisplaybreaks

\newcommand{\beq}{\begin{equation}}

\newcommand{\eeq}{\end{equation}}

 \newcommand{\be}{\begin{equation}}
 \newcommand{\ee}{\end{equation}}
 \newcommand{\bea}{\begin{eqnarray}}
 \newcommand{\eea}{\end{eqnarray}}

\usepackage{tikz}
\usetikzlibrary{positioning}
\usetikzlibrary{intersections}
\usetikzlibrary{fadings} 
\usetikzlibrary{arrows.meta} 
\usetikzlibrary{arrows}

\tikzfading[name=fade out,
inner color=transparent!0,
outer color=transparent!100]

\definecolor{cherryblossompink}{rgb}{1.0, 0.72, 0.77}
\definecolor{lightblue}{rgb}{0.68, 0.85, 0.9}

\usetikzlibrary{decorations.pathmorphing}
\usetikzlibrary{decorations.pathreplacing,decorations.markings}

\usetikzlibrary{backgrounds,automata}

\title{Action integrals for quantum BTZ black holes}

\author{Yuanfan Cao}
\author{and Andrew Svesko}

\affiliation{Department of Mathematics, King’s College London,
Strand, London, WC2R 2LS, United Kingdom}

\emailAdd{yuanfan.cao@kcl.ac.uk}
\emailAdd{andrew.svesko@kcl.ac.uk}

\abstract{Black holes exactly incorporating quantum matter backreaction effects, namely, quantum black holes, are notoriously difficult to construct, let alone study their horizon thermodynamics. Here, we derive the thermodynamics of three-dimensional charged and rotating quantum black holes via the tree-level gravitational partition function. Specifically, we primarily focus on holographic quantum BTZ black holes, dual to $(3+1)$-dimensional accelerating black holes in anti-de Sitter space that localize on Karch-Randall end-of-the-world (ETW) branes. To derive their horizon thermodynamics, we regulate the bulk Euclidean geometry by adding a second ETW brane at asymptotic spatial infinity. We compute the on-shell action of the complexified accelerating black hole in the grand canonical ensemble and derive the quantum BTZ black hole thermodynamics, where the thermal entropy is equal to the generalized entropy. This provides a first principles derivation of the generalized entropy of three-dimensional quantum black holes. Further, we construct charged and rotating quantum black holes in three-dimensional de Sitter and Minkowski space using Randall-Sundrum ETW branes, and compute their horizon thermodynamics.}

\begin{document}

\maketitle

\section{Introduction} \label{sec:intro}

The fact black holes radiate imply they have a thermal description  \cite{Bekenstein:1972tm,Bekenstein:1973ur,Hawking:1974sw,Hawking:1976de}.
Hawking originally showed black holes radiate  
by analyzing quantum matter fields in a fixed black hole background. With this insight, the first law of black hole mechanics \cite{Bardeen:1973gs} is reinterpreted as the first law of black hole thermodynamics, where the Arnowitt–Deser–Misner (ADM) mass is identified as internal energy, and the entropy is proportional to the cross-sectional area of the event horizon.  A first principles account of black hole thermodynamics comes via the Gibbons-Hawking prescription \cite{Gibbons:1976ue} for the gravitational partition function. The ADM mass as internal energy and Bekenstein-Hawking area-entropy are then derived from evaluating the partition function perturbatively about solutions to the classical field equations.

Black hole evaporation will occur 
when the quantum matter fields modeling radiation backreact on the classical dynamical geometry. In an appropriate limit, such dynamics are partially characterized by the semi-classical Einstein equations,
\beq G_{ab}(g)+\Lambda g_{ab}=8\pi G_{N}\langle T_{ab}^{\text{mat}}\rangle\;.\label{eq:semieineq}\eeq
Here $\langle T_{ab}^{\text{mat}}\rangle$ is the expectation value of the (renormalized)  stress-energy tensor of the quantum matter in an appropriate quantum state. Semi-classical gravity should be viewed as an approximation to a more fundamental theory, valid only in a certain regime. Indeed, the semi-classical approximation certainly fails near the Planck scale, and the field equations (\ref{eq:semieineq}) are not expected to be valid for generic quantum states (e.g., macroscopic superpositions \cite{Page:1981aj}). Above the Planck scale, moreover, the semi-classical approximation is not generically self-consistent: metric and matter fluctuations enter at the same order such that it is inconsistent to treat the background geometry classically \cite{Ford:1982wu}. Finally, black hole solutions to the semi-classical field equations, namely, ``quantum black holes'', are technically difficult to uncover. Solving (\ref{eq:semieineq}) amounts to solving a coupled system of geometry and renormalized quantum correlators. A straightforward approach is to treat backreaction perturbatively, yielding an iterative procedure which typically cannot be solved beyond leading quantum corrections, cf. \cite{York:1983zb,York:1984wp,Hochberg:1992rd,Hochberg:1992xt,Anderson:1994hh,Souradeep:1992ia,Steif:1993zv,Matschull:1998rv,Casals:2016ioo,Casals:2019jfo}. Thus, finding \emph{exact} quantum black holes remains a largely open problem.

 A framework in which the semi-classical approximation can be consistently realized is in braneworld holography \cite{deHaro:2000wj}. In this context, a codimension-1 end-of-the-world (ETW) brane is inserted into a bulk $(d+1)$-dimensional anti-de Sitter (AdS) spacetime which is assumed to have a dual description in terms of a $d$-dimensional conformal field theory (CFT) \emph{\'{a} la} AdS/CFT. The boundary conditions are chosen such that the brane geometry is dynamical and, from the brane perspective, the (induced) theory is interpreted as an effective semi-classical theory of gravity with higher-derivative corrections incorporating backreaction effects due to the CFT living on the brane. This interpretation is valid in the large central charge $c$, planar diagram limit of the CFT, such that the bulk theory is governed by Einstein-Hilbert gravity, while the semi-classical brane gravity theory is consistently understood as an asymptotic expansion in $cG_{N}\hbar$, with metric fluctuations suppressed relative to matter loops as $G_{N}\hbar\to0$. Advantageously, via this set-up, quantum black holes can be exactly constructed in $(2+1)$-dimensions and, in principle, beyond, by searching for \emph{classical} bulk black holes that localize on ETW branes \cite{Emparan:2002px}. See \cite{Panella:2024sor} for a review on such constructions.\footnote{A non-holographic setting in which quantum black holes can be consistently and exactly uncovered are in  models of $(1+1)$-dimensional dilaton-gravity coupled to conformal matter, e.g., \cite{Christensen:1977jc,Callan:1992rs,Russo:1992ax,Bose:1995pz,Fabbri:1995bz,Fabbri:2005mw}.}

In this article, we derive the semi-classical thermodynamics of (2+1)-dimensional charged and rotating quantum black holes via the Gibbons-Hawking prescription. The computational strategy is, in principle, straightforward, as we need only evaluate the on-shell Euclidean action of a family of AdS$_{4}$ braneworld black holes. What makes this evaluation technically non-trivial is that the family of bulk black holes are charged and rotating AdS$_{4}$ C-metrics, i.e., charged and rotating black holes that also accelerate in asymptotically AdS$_{4}$ space \cite{Kinnersley:1970zw,Podolsky:2002nk,Podolsky:2006px}. The acceleration is generated due to a cosmic string emanating from the conformal boundary attached at the horizon, pulling the black hole away from the AdS$_{4}$ center (see Figure \ref{fig:accelbh}). The cosmic string distorts the horizon geometry, producing conical singularities, which typically cannot all be removed simultaneously.\footnote{An exception to this are the class of supersymmetric AdS$_{4}$ C-metrics which can be uplifted to 11-dimensional Sasaki-Einstein spacetimes free of conical singularities \cite{Cassani:2021dwa}.} The presence of the cosmic string and an acceleration horizon makes deriving and understanding the thermodynamics much more complicated than non-accelerating, stationary black holes (for progress, see \cite{Appels:2016uha,Appels:2017xoe,Appels:2018jcs,Anabalon:2018ydc,Ball:2020vzo,Arenas-Henriquez:2023hur}). 

Here we take advantage of the braneworld construction to compute the on-shell Euclidean action of accelerating black holes and, subsequently, three-dimensional quantum black holes. The key insight is that the C-metric has two umbilic surfaces, codimension-1 hypersurfaces whose  extrinsic curvature is proportional to the induced (brane) metric. Such surfaces are natural locations to place ETW branes as the Israel junction conditions -- boundary conditions for the variational problem to be well-posed -- are automatically satisfied. Typically, only one such umbilic surface/ETW brane is employed (along a fixed angular coordinate), where, in the bulk Lorentzian AdS$_{4}$ picture, the cosmic string is removed and the black hole sticks to the brane. This brane alone is used to remove all but one of the conical singularities at the horizon. We then place an ETW brane at the second umbilic surface (along a fixed bulk radial position) which serves as an infrared regulator in the Euclideanized spacetime, such that the bulk on-shell action is finite (without employing any ambiguous background subtraction or local counterterms). Our strategy extends a method initiated by Kudoh and Kurita to examine the thermodynamics of neutral and static braneworld black holes \cite{Kudoh:2004ub}. 

Using this set-up, we specifically derive the thermodynamics of ``quantum BTZ'' black holes \cite{Emparan:2020znc,Climent:2024nuj,Feng:2024uia,Bhattacharya:2025tdn} (so named on account of their relation to the classical Ba\~{n}ados-Teteilboim-Zanelli (BTZ) black hole \cite{Banados:1992wn,Banados:1992gq} modified due to perturbative backreaction of a conformally coupled scalar field \cite{Souradeep:1992ia,Steif:1993zv,Matschull:1998rv,Casals:2016ioo,Casals:2019jfo}). In particular, we will show the mass of the charged and rotating AdS$_{4}$ C-metric is equal to the internal energy of the quantum BTZ black hole, and that the four-dimensional classical Bekenstein-Hawking entropy $S_{\text{BH}}^{(4)}$ is equal to the three-dimensional \emph{generalized} entropy \cite{Emparan:2020znc}.
Both will be derived via the bulk canonical partition function. Thus, via a semi-classical saddle-point approximation, we derive the generalized entropy of the quantum black hole from an on-shell Euclidean action,
\beq (\beta \partial_{\beta}-1)I_{E}^{\text{on-shell}}(\beta)=S^{(3)}_{\text{gen}}=S^{(3)}_{\text{BH}}+\nu S^{(3)}_{\text{CFT}}+\nu^{2} S^{(3)}_{\text{Iyer-Wald}}+...\;,\label{eq:entintro}\eeq
for inverse temperature $\beta$.
In practice, we evaluate the on-shell (Euclidean) action $I_{E}^{\text{on-shell}}$ of the regulated AdS$_{4}$ C-metric. Via ``double holography'', however, the classical bulk action is equivalent to the on-shell Euclidean action of the induced brane theory, including contributions from both the gravitational and CFT matter sectors. We emphasize that the first equality is an exact relation, valid for any strength in backreaction, even non-perturbatively. Meanwhile, the second equality follows from a perturbative expansion in semi-classical backreaction, controlled by the dimensionless parameter $\nu \equiv G_{3}\hbar c/\ell_{3}<1$, for AdS$_{3}$ length scale $\ell_{3}$. The generalized entropy decomposes into the three-dimensional Bekenstein-Hawking entropy, the fine grained entropy of the CFT, and the Iyer-Wald entropy due to the leading higher-derivative corrections; the ellipsis represents contributions to the generalized entropy that mixes  higher-derivative corrections in the gravitational and matter sectors at higher orders in $\nu$. Our work provides a first principles derivation of the generalized entropy of exact three-dimensional quantum black holes.\footnote{A first principles derivation of generalized entropy of non-holographic (1+1)-dimensional quantum black holes can be accomplished via the on-shell ``microcanonical'' action \cite{Pedraza:2021ssc,Svesko:2022txo,Alexandre:2025hkr}.}

The remainder of this article is as follows. In Section \ref{sec:qBTZbhs} we review the construction of charged and rotating  quantum BTZ (qBTZ) black holes, where we explicitly compute the quantum matter stress-tensor supporting the geometry. In Section \ref{sec:qBHthermo} we derive the thermodynamics of quantum BTZ black holes via the grand canonical partition function. Using Randall-Sundrum ETW branes, in Section \ref{sec:bhsindSMink3} we construct quantum black holes in $(2+1)$-dimensional de Sitter and Minkowski spacetimes that are both charged and rotating, generalizing \cite{Emparan:2022ijy,Panella:2023lsi,Panella:2024sor,Climent:2024wol}, and evaluate their horizon thermodynamics. We conclude in Section \ref{sec:disc}. To keep this article self-contained, we include Appendix \ref{app:scgravrev}, reviewing the self-consistency of semi-classical gravity and the induced braneworld theory, and Appendix \ref{app:cmetgeom}, providing additional details about the geometry of the AdS$_{4}$ C-metric. Unless otherwise stated, we work in units where the speed of light and Planck's constant $\hbar$ are set to unity.


\section{Quantum BTZ black holes} \label{sec:qBTZbhs}

Here we review the holographic construction of quantum BTZ black holes.

\subsection{Set-up: bulk and brane}

All known exact constructions of three-dimensional quantum black holes are understood to correspond to specific asymptotically AdS$_{4}$ black holes that localize on an end-of-the-world brane \cite{Emparan:1999wa,Emparan:1999fd,Emparan:2002px} (see \cite{Panella:2024sor} for a review). We take the bulk to be characterized by Einstein-Maxwell-AdS$_{4}$ gravity, 
\beq I=\frac{1}{16\pi G_{4}}\int d^{4}x\sqrt{-\hat{g}}\left[\hat{R}+\frac{6}{\ell_{4}^{2}}-\frac{\ell_{\star}^{2}}{4}\hat{F}^{2}\right]\;, \quad \ell^{2}_{\star}=\frac{16\pi G_{4}}{g_{\star}^{2}}\;.\eeq
Here  $\hat{g}_{ab}$ denotes the bulk asymptotically AdS$_{4}$ metric with Ricci scalar $\hat{R}$, $G_{4}$ is the four-dimensional Newton's constant, $\hat{F}_{ab}=\partial_{a}A_{b}-\partial_{b}A_{a}$ is the bulk Maxwell field strength tensor, $\ell_{\star}$ is a coupling constant with dimensions of length, $g_{\star}$ is the dimensionless gauge coupling constant, and $\ell_{4}$ is the AdS$_{4}$ radius. For the theory to have a well-posed variational problem, we append the bulk action with a Gibbons-Hawking-York boundary term (suitable for Dirichlet boundary conditions). We will return to this point in Section \ref{sec:qBHthermo}. 

We minimally couple the bulk theory to a codimension-1 ETW brane $\mathcal{B}$ of tension $\tau$ characterized by, for simplicity, a purely tensional action 
\beq I_{\text{brane}}=-\tau\int_{\mathcal{B}}d^{3}x\sqrt{-h}\;.\label{eq:genbraneact}\eeq
Here $h_{ij}$ is the (induced) metric on the brane. For the present purposes, the brane $\mathcal{B}$ is a Karch-Randall brane \cite{Karch:2000ct,Karch:2000gx}, taken to have AdS$_{3}$ asymptotics. 
As an ETW brane, a small portion of the AdS$_{4}$ bulk, including its conformal boundary, is cut-off. We work with a $\mathbb{Z}_{2}$-construction by surgically completing the space by introducing a second copy of the AdS$_{4}$ bulk plus brane geometry and suturing along the common $\mathcal{B}$.
This procedure leads to a jump discontinuity in the extrinsic curvature $K_{ij}$ across the junction, characterized by the Israel junction conditions \cite{Israel:1966rt},
\beq \Delta K_{ij}-h_{ij}\Delta K=-8\pi G_{4}S_{ij}=8\pi G_{4}\tau h_{ij}\;.\label{eq:israeljuncconds}\eeq
Here $\Delta K_{ij}=K^{+}_{ij}-K_{ij}^{-}$ denotes the difference between the extrinsic curvature across either `$+$' and `$-$' sides of the brane,\footnote{Here we take $K_{ij}^{+}=-K_{ij}^{-}$ such that $\Delta K_{ij}=2K_{ij}$.} and $S_{ij}$ is the brane stress-tensor; for a purely tensional brane (\ref{eq:genbraneact}) one has $S_{ij}\equiv-\frac{2}{\sqrt{-h}}\frac{\delta I_{\text{brane}}}{\delta h^{ij}}=-\tau h_{ij}$.  
The bulk Maxwell field strength must also obey junction conditions \cite{Lemos:2021jtm},
\beq 
\begin{split}
&\Delta \hat{F}_{ij}=\hat{F}^{+}_{ij}-\hat{F}^{-}_{ij}=0\;,\\
&\Delta f_{i}=f^{+}_{i}-f^{-}_{i}=4\pi j_{i}\;,
\end{split}
\label{eq:junccondEM}\eeq
where $\hat{F}_{ij}\equiv \hat{F}_{ab}e^{a}_{i}e^{b}_{j}$ and $f_{i}\equiv \hat{F}_{ab}e^{a}_{i}n^{b}$ are projected components of $\hat{F}_{ab}$ onto the brane
and $j_{i}$ is the electromagnetic surface current (for a basis $e^{a}_{i}$ of vectors tangent to the brane and $n^{a}$ is unit and normal to the brane). 

\vspace{2mm}

\begin{figure}[t!]
\centering
 \includegraphics[width=5.5cm]{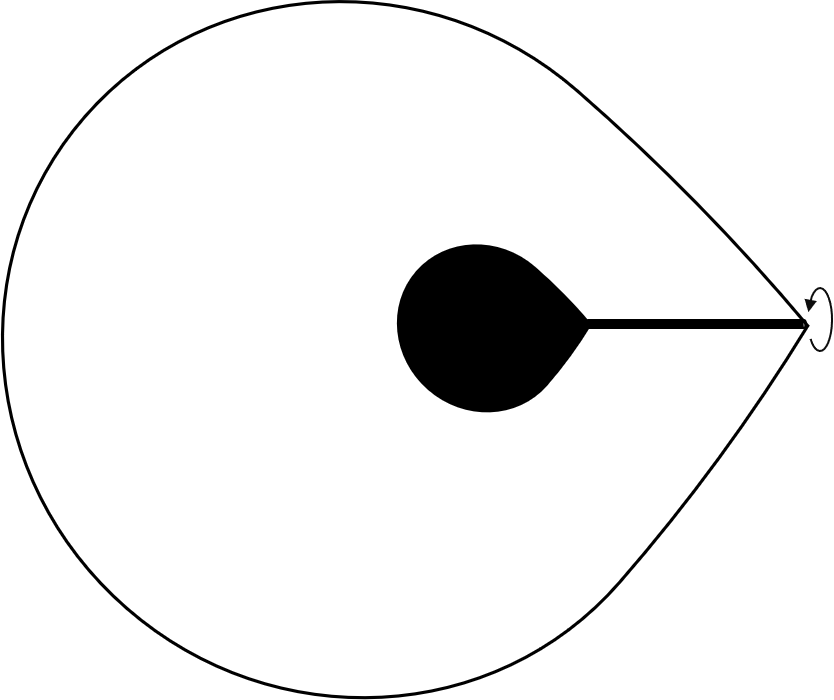}
\put(-1,87){ $\phi$}
\caption{\small \textbf{Accelerating black hole in AdS$_{4}$}. A constant $t$ and $\phi$ slice of the $\text{AdS}_{4}$ C-metric. Conical singularities distort the black hole horizon, giving it a conical shape at one pole where a cosmic string is attached, suspending the black hole away from the center.}
\label{fig:accelbh}\end{figure}

\noindent \textbf{AdS$_{4}$ C-metric.} We are interested in stationary AdS$_{4}$ black holes which localize on the brane $\mathcal{B}$. With respect to the bulk, a brane with non-vanishing tension is an accelerated trajectory. Thus, a black hole that localizes on the brane is in an accelerated frame. A natural candidate black hole solution to the Einstein-Maxwell field equations coupled to a brane, then, is the AdS$_{4}$ C-metric, which may be interpreted as a black hole uniformly accelerating in  AdS$_{4}$ due to a cosmic string attached to the horizon \cite{Griffiths:2006tk}, see Figure \ref{fig:accelbh} for an illustration. 

The metric for a charged and rotating AdS$_{4}$ C-metric in Boyer-Lindquist coordinates is
\beq
\begin{split}
 ds^2= \frac{\ell^2}{(\ell + xr)^2} \biggr[ & -\frac{H(r)}{\Sigma(x,r)} \left(dt+ax^2 d\phi \right)^2 + \frac{\Sigma(x,r)}{H(r)} dr^2 \\
 &+ r^2 \left(\frac{\Sigma(x,r)}{G(x)}dx^2 + \frac{G(x)}{\Sigma(x,r)}\left( d\phi- \frac{a}{r^2} dt \right)^2  \right) \biggr] \ ,
 \end{split}
\label{eq:rotatingCmetqbtz}
\eeq
with metric functions
\beq
\begin{split}
    &H(r)= \frac{r^2}{\ell_3^2}+ \kappa -\frac{\mu \ell}{r}+ \frac{a^2}{r^2}+\frac{q^{2}\ell^{2}}{r^{2}}  \ , \quad G(x)= 1-\kappa x^2-\mu x^3+ \frac{a^2}{\ell_3^2}x^4-q^{2}x^{4} \ , \\
   &\Sigma(x,r)= 1 + \frac{a^2 x^2}{r^2} \ ,
\end{split}
\label{eq:metfuncsrotqbtz}\eeq
and AdS$_{4}$ length scale
\beq \frac{1}{\ell_{4}^{2}}=\frac{1}{\ell_{3}^{2}}+\frac{1}{\ell^{2}}\;.\label{eq:AdS4lenghscale}\eeq
Here $\kappa=\pm1,0$ corresponds to types of slicings of the boundary, $\mu$ is a dimensionless parameter related to the mass of the black hole, the non-negative parameter $a$ (of dimensions length) is related to the angular rotation, and $q$ serves as a dimensionless (electric) charge parameter. Finally, $\ell$ is a length equal to the inverse acceleration of the black hole and $\ell_{3}$ is the length scale of the AdS$_{3}$ brane geometry. To keep the AdS$_{4}$ length scale (\ref{eq:AdS4lenghscale}) positive, we require 
\beq \ell_{4}<\ell_{3}\;,\label{eq:restricads4l}\eeq
which is valid for all $0\leq \ell<\infty$.

The asymptotic AdS$_{4}$ (timelike) boundary is located where the conformal factor in the metric (\ref{eq:rotatingCmetqbtz}) vanishes, i.e., at $xr=-\ell$. For $\mu\neq0$, the metric (\ref{eq:rotatingCmetqbtz}) has inner and outer black hole horizons, $r_{-}$ and $r_{+}$, respectively, satisfying $H(r_{\pm})=0$ and $0<r_{-}<r_{+}$. The geometry has an extremal limit when the inner and outer horizons coincide, $r_{-}=r_{+}$. The upper bound on the AdS$_{4}$ length (\ref{eq:restricads4l}) pushes the non-compact Rindler-like acceleration horizon beyond our regime of interest, such that $H(r)$ will only have positive roots $r_{\pm}$. Specifically, rearranging (\ref{eq:restricads4l}), the metric function $H(r)$ is 
\beq H(r)=\kappa+\frac{r^{2}}{\ell^{2}_{4}}-\frac{r^{2}}{\ell^{2}}-\frac{\mu\ell}{r}+\frac{a^{2}}{r^{2}}+\frac{q^{2}\ell^{2}}{r^{2}}\;.\eeq
We see the acceleration and negative curvature of AdS counteract one another such that the restriction (\ref{eq:restricads4l}), forcing $\ell>\ell_{4}$, effectively removes the acceleration horizon. Due to this feature, the AdS$_{4}$ C-metric is said to be slowly accelerating \cite{Podolsky:2002nk}.

The real zeros $x_{i}$ of $G(x)$ characterize the shape of the horizon, corresponding to symmetry axes of the Killing vector $\partial^{a}_{\phi}$. Each distinct zero gives rise to a distinct conical singularity on the horizon.  One of these conical singularities can be removed by imposing regularity on the azimuthal coordinate $\phi$ to ensure smoothness of the geometry along the axis of rotational symmetry, 
\beq \phi\sim \phi+\Delta\phi\;,\quad \Delta\phi=\frac{4\pi}{|G'(x_{i})|}=4\pi\biggr|-2\kappa x_{i}-3\mu x_{i}^{2}-4\left(q^{2}-\frac{a^{2}}{\ell^{2}_{3}}\right)x_{i}^{3}\biggr|^{-1}\;.\label{eq:azimuthid}\eeq
Denote the smallest positive root as $x=x_{1}$. We fix the periodicity such that the conical singularity at $x_{i}=x_{1}$ is removed, leaving conical singularities at zeroes $x_{i}\neq x_{1}$. The restricted region $0\leq x\leq x_{1}$ is regular and free of conical singularities. Clearly,  $x_{1}$ is a function of $\mu,q$ and $a$. Instead, for convenience, we view $\mu$ as a `derived' parameter by solving $G(x_{1})=0$, 
\beq \mu=\frac{1}{x_{1}^{3}}\left[-\left(q^{2}-\frac{a^{2}}{\ell_{3}^{2}}\right)x_{1}^{4}-\kappa x_{1}^{2}+1\right]\;,\label{eq:muderiv}\eeq
which we take to be non-negative.

Finally, the charged C-metric (\ref{eq:rotatingCmetqbtz}) is supported by the gauge field 1-form, 
\beq A=A_{b}dx^{b}=-\frac{2}{\ell_{\star}}\frac{\ell qr}{r^{2}+a^{2}x^{2}}dt-\frac{2}{\ell_{\star}}\frac{a\ell qrx^{2}}{r^{2}+a^{2}x^{2}}d\phi\;,\label{eq:gaugefieldBL}\eeq
with field strength $\hat{F}=dA$.
Together with the metric, it is straightforward to verify the AdS$_{4}$ C-metric is a solution to Einstein's equations
\beq \hat{R}_{ab}+\frac{\ell^{2}_{\star}}{2}\hat{F}_{a}^{\;c}\hat{F}_{cb}+\frac{\ell^{2}_{\star}}{8}\hat{g}_{ab}\hat{F}_{cd}^{2}=-\frac{3}{\ell_{4}^{2}}\hat{g}_{ab}\;,\eeq
and Maxwell's equations $\nabla_{a}\hat{F}^{ab}=0$. 

\vspace{2mm}

\noindent \textbf{Umbilic surfaces and ETW branes.} An advantageous feature of the C-metric (\ref{eq:rotatingCmetqbtz}) is that the hypersurface $x=0$ is totally umbilic, i.e., the extrinsic curvature $K_{ij}$ is proportional to the induced metric at $x=0$. In particular,  $K_{ij}=-\ell^{-1}h_{ij}$. Thus, a codimension-1 brane $\mathcal{B}$ placed at $x=0$ is guaranteed to obey the Israel junction conditions (\ref{eq:israeljuncconds}), from which the tension is readily identified to be
\beq \tau=\frac{1}{2\pi G_{4}\ell}\;.\label{eq:branetengen}\eeq
Notice that in the limit of vanishing acceleration, $\ell\to\infty$, one has $\ell_{4}\to\ell_{3}$ and the brane becomes tensionless. This is a special case of an umbilic boundary, said to be geodesic, having vanishing geodesic curvature intrinsic to the surface. For $\ell$ non-zero and finite, the brane undergoes non-geodesic motion. Meanwhile, in the opposite limit, $\ell\to0$, the brane approaches the AdS$_{4}$ conformal boundary and $\ell_{4}\to\ell$. In this limit, upon a double-Wick rotation, the geometry at the boundary is that of a (non-dynamical) rotating and charged defect in AdS$_{3}$; for vanishing charge the boundary is the rotating BTZ black hole \cite{Hubeny:2009rc} (see Appendix B of \cite{Emparan:2020znc} for the appropriate double-Wick rotation), while for vanishing rotation one has a charged defect in AdS$_{3}$ \cite{Climent:2024nuj}. 

Treating the $x=0$ hypersurface as an ETW brane, the region $x<0$ (where the remaining conical singularities reside) is cutoff from the rest of bulk AdS$_{4}$ space.  We then complete the space by introducing a second copy of the remaining $0\leq x\leq x_{1}$ region, glued along the common $x=0$ hypersurface. This surgical procedure results in a $\mathbb{Z}_{2}$-symmetric double-sided braneworld \cite{Randall:1999vf,Karch:2000ct} free of conical singularities.

\vspace{2mm}

\noindent \textbf{Localized black holes on the brane.} The bulk AdS$_{4}$ black hole (\ref{eq:rotatingCmetqbtz}) will localize on the brane at $x=0$, with induced geometry 
\beq ds^{2}|_{x=0}=-H(r)dt^{2}+H^{-1}(r)dr^{2}+r^{2}\left(d\phi-\frac{a}{r}dt\right)^{2}\;,\label{eq:naivemetrot}\eeq
for metric function $H(r)$ given in (\ref{eq:metfuncsrotqbtz}). The gauge field (\ref{eq:gaugefieldBL}) at $x=0$ is
\beq A_{\mu}dx^{\mu}|_{x=0}=-\frac{2}{\ell_{\star}}\frac{\ell q}{r}dt\;.\eeq
This, however, gives the mistaken impression that the Maxwell gauge field localizes on the brane in the same way as gravity (see \cite{Climent:2024nuj} and Appendix \ref{app:scgravrev}). 

Due to the azimuthal identification (\ref{eq:azimuthid}), the metric (\ref{eq:naivemetrot}) is not cast in canonically normalized coordinates, dubbed the `naive' metric. For the azimuthal coordinate to have range $\phi\in[0,2\pi]$ and for the geometry to take the appropriate form of rotating AdS$_{3}$ as $r\to\infty$, both the $t$ and $\phi$ coordinates must be transformed. We thus introduce canonically normalized coordinates $(\bar{t},\bar{r},\bar{\phi})$ via \cite{Emparan:2020znc} (see also the discussion below Eq. (3.8) of \cite{Panella:2023lsi})
\beq t=\eta (\bar{t}-\bar{a}\ell_{3}\bar{\phi})\;,\quad r=\sqrt{\frac{\bar{r}^{2}-r_{s}^{2}}{(1-\bar{a}^{2})\eta^{2}}}\;,\quad \phi=\eta\left(\bar{\phi}-\frac{\bar{a}\bar{t}}{\ell_{3}}\right)\;,\label{eq:transKillvec}\eeq
where 
\beq \eta\equiv \frac{\Delta\phi}{2\pi}\;,\quad \bar{a}\equiv \frac{ax_{1}^{2}}{\ell_{3}}\;,\quad r_{s}\equiv \frac{\ell_{3}\bar{a}\eta}{x_{1}}\sqrt{2-\kappa x_{1}^{2}}\;.\label{eq:transpara}\eeq
In these coordinates, the induced metric on the brane is \cite{Bhattacharya:2025tdn}
\beq 
\begin{split}
ds^{2}|_{x=0}&=-\left[\frac{\bar{r}^{2}}{\ell_{3}^{2}}-\eta^{2}\left(\frac{4\bar{a}^{2}}{x_{1}^{2}}-(1+\bar{a}^{2})\kappa\right)-\frac{\mu\ell\eta^{2}}{r}+\frac{q^{2}\ell^{2}\eta^{2}}{r^{2}}\right]d\bar{t}^{2}\\
&+\frac{\bar{r}^{2}}{(\bar{a}^{2}-1)\eta^{4}r^{2}}\left[\kappa+\frac{r^{2}}{\ell_{3}^{2}}-\frac{\mu\ell}{r}+\frac{q^{2}\ell^{2}x_{1}^{2}+\bar{a}^{2}\ell_{3}^{2}}{x_{1}^{4}r^{2}}\right]^{-1}d\bar{r}^{2}\\
&+\left[\bar{r}^{2}-\bar{a}^{2}\eta^{2}\ell_{3}^{2}\left(\frac{q^{2}\ell^{2}}{r^{2}}-\frac{\mu\ell}{r}\right)\right]d\bar{\phi}^{2}\\
&-2\bar{a}x_{1}\ell_{3}\eta^{2}(\mu+q^{2}x_{1})\left[1+\frac{\ell}{x_{1}}-\frac{q^{2}\ell(x_{1}r+\ell)}{r^{2}x_{1}(q^{2}x_{1}+\mu)}\right]d\bar{t}d\bar{\phi}\;.
\end{split}
\label{eq:cannormmet}\eeq
Here the metric is written with both $r=r(\bar{r})$ and $\bar{r}$, out of convenience. Cast like this, the canonically normalized metric (\ref{eq:cannormmet}) has points identified along orbits of Killing vector $\partial_{\bar{\phi}}$, with $(\bar{t},\bar{\phi})\sim (\bar{t},\bar{\phi}+2\pi)$ as desired.%

The gauge potential (\ref{eq:gaugefieldBL}) in canonically normalized coordinates has the form
\beq
\begin{split} 
\bar{A}_{b}d\bar{x}^{b}&=-\frac{2q\ell r\eta}{\ell_{\star}(r^{2}+a^{2}x^{2})}\left(1-\frac{a^{2}x^{2}x_{1}^{2}}{\ell_{3}^{2}}\right)d\bar{t}-\frac{2q\ell r\eta}{\ell_{\star}(r^{2}+a^{2}x^{2})}(ax^{2}-ax_{1}^{2})d\bar{\phi}\;,
\end{split}
\label{eq:Agaugbarred}\eeq
such that on the brane at $x=0$ it reduces to
\beq \bar{A}_{b}d\bar{x}^{b}|_{x=0}=-\frac{2q\ell\eta}{\ell_{\star}r(\bar{r})}d\bar{t}+\frac{2\bar{a}q\ell \ell_{3}\eta}{\ell_{\star}r(\bar{r})}d\bar{\phi}\;.\label{eq:gaugepotqbtz}\eeq

The ETW brane at $x=0$ intersects the bulk black hole horizon such that the braneworld geometry (\ref{eq:naivemetrot}) or (\ref{eq:cannormmet}) retains the same inner and outer horizons as for the bulk black hole. Indeed, roots $r_{\pm}$ of $H(r)$ correspond to when the generator of the (bulk) Killing horizon 
\beq \zeta^{b}_{\pm}=\partial^{b}_{t}+\frac{a}{r_{\pm}^{2}}\partial^{b}_{\phi}\;,\eeq
become null. Via the transformation (\ref{eq:transKillvec}), the Killing vectors $\partial_{t}$ and $\partial_{\phi}$ transform as \cite{Emparan:2020znc,Panella:2024sor}
\beq \partial_{t}^{b}=\frac{1}{\eta(1-\bar{a}^{2})}\left(\partial^{b}_{\bar{t}}+\frac{\bar{a}}{\ell_{3}}\partial^{b}_{\bar{\phi}}\right)\;,\quad \partial_{\phi}^{b}=\frac{1}{\eta(1-\bar{a}^{2})}\left(\partial_{\bar{\phi}}^{b}+\bar{a}\ell_{3}\partial_{\bar{t}}^{b}\right)\;.\label{eq:killvec}\eeq
In canonically normalized coordinates 
the generator of the Killing horizon $\bar{\zeta}^{b}$ is 
\beq \bar{\zeta}^{b}_{\pm}=\partial^{b}_{\bar{t}}+\bar{\Omega}_{\pm}\partial^{b}_{\bar{\phi}}\;,\label{eq:cannormgenerator}\eeq
    for uniform (inner/outer) horizon angular velocity as measured at spatial infinity\footnote{In the tensionless limit, $\ell\to\infty$, the charged and rotating AdS$_{4}$ C-metric is known to reduce to the Kerr-Newman-AdS$_{4}$ black hole with $\ell_{3}=\ell_{4}$, after taking appropriate coordinate and parameter rescalings (see, e.g., Appendix B of \cite{Bhattacharya:2025tdn}). In particular, for $a=\hat{a}/x_{1}^{2}$ and $r=\hat{r}/x_{1}$, the angular velocity (\ref{eq:horangvel}) is equivalent to that for Kerr-Newman-AdS$_{4}$, cf. Eq. (17) of \cite{Caldarelli:1999xj}. \label{fn:tenlimit}} 
\beq \bar{\Omega}_{\pm}=\frac{a}{r_{\pm}^{2}+a^{2}x_{1}^{2}}\left(1+\frac{r_{\pm}^{2}x_{1}^{2}}{\ell_{3}^{2}}\right)\;.\label{eq:horangvel}\eeq
Inverting (\ref{eq:killvec}), it is easy to show 
\beq \bar{\zeta}^{b}_{\pm}=\frac{\eta(1-\bar{a}^{2})}{1+\frac{a^{2}x_{1}^{2}}{r_{\pm}^{2}}}\zeta^{b}_{\pm}\;,\eeq
such that $\bar{\zeta}_{\pm}$ and $\zeta_{\pm}$ generate the same horizon.  The co-rotating angular velocities measured at the horizon, $\bar{\Omega}^{\text{co}}_{+}$, and spatial infinity, $\bar{\Omega}^{\text{co}}_{\infty}$ are,
\beq \bar{\Omega}_{+}^{\text{co}}=\frac{a}{r_{+}^{2}+a^{2}x_{1}^{2}}(1-\bar{a}^{2})\;,\quad \bar{\Omega}^{\text{co}}_{\infty}=-\frac{\bar{a}}{\ell_{3}}\;,\eeq
such that $\bar{\Omega}_{+}=\bar{\Omega}^{\text{co}}_{+}-\bar{\Omega}^{\text{co}}_{\infty}$.

Relative to the canonically normalized horizon generator (\ref{eq:cannormgenerator}), the surface gravity $\bar{\kappa}$ is defined as $\nabla_{b}(-\bar{\zeta}^{2})\equiv 2\bar{\kappa}\bar{\zeta}_{b}$. To evaluate the surface gravity at the horizon, define $\bar{\zeta}^{2}\equiv -\lambda^{2}$ (for constant function $\lambda$) such that $\nabla_{b}(\lambda^{2})=2\lambda\nabla_{b}\lambda=2\lambda \partial_{\nu}\lambda$, and  $\bar{\kappa}^{2}=-g^{ab}(\partial_{a}\lambda)(\partial_{b}\lambda)$. Evaluating this at the inner/outer horizon, we find
\beq \bar{\kappa}_{\pm}=\frac{\eta(1-\bar{a}^{2})}{1+\frac{a^{2}x_{1}^{2}}{r_{\pm}^{2}}}\frac{H'(r_{\pm})}{2}=\frac{\eta(2r_{\pm}^{4}-2\ell_{3}^{2}(a^{2}+q^{2}\ell^{2})+\mu r_{+}\ell_{3}^{2}\ell)}{2r_{\pm}\ell_{3}^{2}(a^{2}x_{1}^{2}+r_{\pm}^{2})}\;,\label{eq:surfacegrav}\eeq
consistent with Eq. (2.40) of \cite{Bhattacharya:2025tdn}.

\subsection{Quantum black holes on the brane}

Thus far we have taken the bulk perspective. Let us now apply the formalism of double holography to extract the viewpoint of a being confined to the brane. 

\vspace{2mm} 

\noindent \textbf{Karch-Randall braneworld holography.} Integrating out the bulk degrees of freedom as done in holographic renormalization \cite{deHaro:2000vlm,Skenderis:2002wp} (see also \cite{Emparan:2023dxm,Panella:2024sor}) leaves an effective three-dimensional theory of gravity on the brane coupled to a CFT$_{3}$ with central charge $c_{3}$ and an ultraviolet cutoff $\ell$. Schematically, 
\beq I=I_{\text{Bgrav}}[\mathcal{B}]+I_{\text{CFT}}[\mathcal{B}]\;.\label{eq:genactind}\eeq
The induced gravity theory characterized by action $I_{\text{Bgrav}}$ is a specific higher-derivative theory of gravity, including an infinite tower of higher-curvature terms and non-minimal coupling with the Maxwell field. The precise form of the brane-induced action will not be necessary for our purposes (see \cite{Climent:2024nuj,Panella:2024sor} and Appendix \ref{app:scgravrev} for further details). Relevant are the effective brane couplings induced from the couplings characterizing the four-dimensional bulk theory:
\bea
G_{3}&=&\frac{G_{4}}{2\ell_{4}} \,,\label{eq:effGd}\\
\frac{1}{L_{3}^2}&=&\frac{2}{\ell_{4}^2}\left(1-2\pi G_{4}\ell_{4}\tau\right)=\frac{2}{\ell_{4}^{2}}\left(1-\frac{\ell_{4}}{\ell}\right)\,,\\
\tilde{\ell}_{\star}^{2}&=&\frac{5}{4}\ell_{\star}^{2}\;,\quad \tilde{g}_{\star}^{2}=\frac{2}{5}\frac{g_{\star}^{2}}{\ell_{4}}\;.
\label{eq:effLd}
\eea
Here $L_3$ is the effective AdS$_{3}$ radius on the brane such that the induced cosmological constant $\Lambda_{3}=-1/L_{3}^{2}$. 
The induced electromagnetic couplings $\tilde{\ell}_{\star}$ and $\tilde{g}_{\star}$ are such that $\tilde{\ell}_{\star}^{2}=\frac{16\pi G_{3}}{\tilde{g}_{\star}^{2}}$. 

Generally, the three-dimensional gravity theory is massive. The brane graviton, however, will become effectively massless in the limit the brane is near the boundary. Using the bulk length scale (\ref{eq:AdS4lenghscale}), it follows in this limit
\beq \frac{1}{L_{3}^{2}}=\frac{1}{\ell_{3}^{2}}\left[1+\frac{\ell^{2}}{4\ell_{3}^{2}}+\mathcal{O}\left(\frac{\ell^{4}}{\ell_{3}^{4}}+...\right)\right]\;,\label{eq:L3sqrel}\eeq
with $\ell\sim \ell_{4}\ll \ell_{3}$. Thus, to leading order $L_{3}\approx \ell_{3}$. Further note the central charge $c_{3}$ of the $\text{CFT}_{3}$ in terms of the  induced brane couplings is
\beq c_{3}=\frac{\ell_{4}^{2}}{G_{4}}=\frac{\ell}{2G_{3}\sqrt{1+(\ell/\ell_{3})^{2}}}\;.\eeq
 Expanding for small $(\ell/\ell_{3})$, i.e., when the brane is near the boundary, gives
\beq 2c_{3}G_{3}=L_{4}\approx \ell\left(1-\frac{\ell^{2}}{2\ell_{3}^{2}}+\frac{3}{8}\frac{\ell^{4}}{\ell_{3}^{4}}\right)\;,\label{eq:c3G3l}\eeq
and $G_{3}c_{3}\sim \ell$. Together, this analysis reveals that, in the limit the 3D graviton is treated  as effectively massless, the semi-classical theory is an asymptotic expansion in small but finite $c_{3}G_{3}\sim \ell$ (with $c_{3}$ assumed to be large), where higher-derivative corrections to 3D Einstein-Hilbert gravity enter at order $\mathcal{O}(\ell^{2})$, while the CFT action enters at order $\mathcal{O}(\ell)$.

Varying the brane theory with respect to the induced brane metric $h_{ij}$ results in the semi-classical equations of motion, reported here up to order $\mathcal{O}(\ell^{2})$ \cite{Climent:2024nuj}
\beq
\begin{split}
&8\pi G_{3}\langle T_{ij}\rangle=R_{ij}-\frac{1}{2}h_{ij}\left(R+\frac{2}{L_{3}^{2}}\right)-\frac{\tilde{\ell}^{2}_{\ast}}{2}\left(F^{k}_{i}F_{jk}-\frac{1}{4}h_{ij}F^{2}\right)+16\pi G_{3} A_{k}j^{k}h_{ij}\\
&+\ell^{2}\biggr[4R_{i}^{\;k}R_{jk}-\frac{9}{4}RR_{ij}-\Box R_{ij}+\frac{1}{4}\nabla_{i}\nabla_{j}R+\frac{1}{2}h_{ij}\left(\frac{13}{8}R^{2}-3R_{kl}^{2}+\frac{1}{2}\Box R\right)\biggr]+\mathcal{O}(\ell^{3})\;,
\end{split}
\label{eq:semiclasseombrane}\eeq
where $\Box\equiv\nabla^{2}$. Meanwhile, varying with respect to the gauge field $A_{i}$ gives the semi-classical Maxwell equations, 
\beq
\begin{split}
\langle J^{j}\rangle&=j^{j}+\frac{\tilde{\ell}^{2}_{\ast}}{16\pi G_{3}}\biggr\{\nabla_{i}F^{ji}+\frac{16}{5}\ell^{2}\biggr(-\frac{1}{72}R\nabla_{i}F^{ji}+\frac{11}{18}F^{j}_{\;\,i}\nabla^{i}R+\frac{209}{294}R^{ij}\nabla_{k}F_{i}^{\;k}\\
&+\frac{5}{4}R^{ik}\nabla_{k}F^{j}_{\;\,i}+\frac{5}{4}F^{ik}\nabla_{k}R^{j}_{\;i}+\frac{317}{588}\nabla_{i}\nabla^{i}\nabla_{k}F^{jk}+\frac{317}{588}\nabla^{j}\nabla^{k}\nabla_{i}F^{ik}\biggr)+\mathcal{O}(\ell^{3})\biggr\}\;.
\end{split}
\label{eq:sccurrentdens}\eeq
Let us expand the quantum stress tensor as $\langle T_{ij}\rangle=\langle T_{ij}\rangle_0+\ell^2 \langle T_{ij}\rangle_2+...$. Then, 
\beq 
    8\pi G_3\langle T\indices{^i_j}\rangle_0=R\indices{^i_j}-\delta \indices{^i_j}\left(R+\frac{2}{\ell^2_3}\right)-\frac{\tilde{\ell}^{2}_{\ast}}{2}\left(F^{ik}F_{jk}-\frac{1}{4}\delta \indices{^i_j}F^{2}\right)+16\pi G_{3} A_{k}j^{k}\delta \indices{^i_j},
\label{eq:Tij0gen}\eeq
\beq 
\begin{split}
    8\pi G_3\langle T\indices{^i_j}\rangle_2&=4R^{ik}R_{jk}-\square R\indices{^i_j}-\frac{9}{4}R R\indices{^i_j}+\frac{1}{4}\nabla^i \nabla_j R\\
    &+\frac{1}{2}\delta \indices{^i_j}\left(\frac{13}{8}R^2-3R^2_{kl}+\frac{1}{2}\square R-\frac{1}{2\ell^4_3}\right)\;.
\end{split}
\label{eq:Tij2gen}\eeq

\vspace{2mm}

\noindent \textbf{Quantum BTZ black hole.} With the induced three-dimensional perspective in mind, let us return to the brane metric (\ref{eq:cannormmet}). From the $\bar{t}\bar{t}$ component, one identifies the black hole mass $M$ by looking at the subleading constant term
\beq 8\mathcal{G}_{3}M=-\kappa\eta^{2}\left(1+\bar{a}^{2}-\frac{4\bar{a}^{2}}{\kappa x_{1}^{2}}\right)\;,\label{eq:massqbtz}\eeq
where $\mathcal{G}_{3}$ denotes a `renormalized' Newton constant due to an all-order resummation of the higher-derivative corrections
to the mass \cite{Emparan:2020znc}; $G_{3}$ represents a ‘bare’ Newton constant. In particular, $\mathcal{G}_{3}\equiv G_{4}/2\ell=G_{3}\ell_{4}/\ell\approx \left(1-\frac{\ell^{2}}{2\ell_{3}^{2}}\right)G_{3}$. That (\ref{eq:massqbtz}) is indeed the mass is further corroborated when evaluating the first law of the thermodynamics, however, at this stage, such an identification is premature.  Meanwhile, the angular momentum $J$ is identified by looking at the $\bar{t}\bar{\phi}$ metric component in the large-$\bar{r}$ limit, 
\beq 4\mathcal{G}_{3}J=\bar{a}x_{1}\ell_{3}\eta^{2}(q^{2}x_{1}+\mu)\;.\label{eq:angmomeqbtz}\eeq

In terms of the mass and angular momentum, the induced metric on the brane (\ref{eq:cannormmet}) may be recast as \cite{Bhattacharya:2025tdn} 
\beq 
\begin{split}
ds^{2}_{\text{qBTZ}}&=-\left[\frac{\bar{r}^{2}}{\ell_{3}^{2}}-8\mathcal{G}_{3}M-\ell\eta^{2}\left(\frac{\mu}{r}-\frac{q^{2}\ell}{r^{2}}\right)\right]d\bar{t}^{2}\\
&+\left[\frac{\bar{r}^{2}}{\ell^{2}_{3}}-8\mathcal{G}_{3}M+\frac{(4\mathcal{G}_{3}J)^{2}}{\bar{r}^{2}}+(\bar{a}^{2}-1)^{2}\ell\eta^{4}\frac{(q^{2}\ell-\mu r)}{\bar{r}^{2}}\right]^{-1}d\bar{r}^{2}\\
&+\left[\bar{r}^{2}+\bar{a}^{2}\ell_{3}^{2}\ell\eta^{2}\left(\frac{\mu}{r}-\frac{q^{2}\ell}{r^{2}}\right)\right]d\bar{\phi}^{2}-8\mathcal{G}_{3}J\left(1+\frac{\ell}{x_{1}r}-\frac{q^{2}\ell(x_{1}r+\ell)}{r^{2}x_{1}(q^{2}x_{1}+\mu)}\right)d\bar{t}d\bar{\phi}\;.
\end{split}
\label{eq:qbtzmet}\eeq
In the vanishing charge limit $q\to0$, one recovers the neutral rotating quantum BTZ (qBTZ) black hole \cite{Emparan:2020znc}. 

In addition to the mass and angular momentum, the electric charge of the black hole is given by the charge as computed in the bulk, 
\beq Q=\frac{2}{g_{\star}^{2}}\int \star\hat{F}=\frac{2q\ell}{g_{\star}^{2}\ell_{\star}}\int_{0}^{2\pi\eta}d\phi\int_{0}^{x_{1}}dx=\frac{8\pi\eta qx_{1}\ell}{g_{\star}^{2}\ell_{\star}}\;,\label{eq:elecchargeqBTZ}\eeq
where the factor of two in front is due to the $\mathbb{Z}_{2}$ symmetry, $\star\hat{F}=r^{2}\hat{F}_{rt}d\phi dx$ refers to the
Hodge dual of the bulk Maxwell tensor, and the integration is taken to be at the boundary. The conjugate electric potential, meanwhile, is given by 
\beq \Phi\equiv \bar{A}_{b}\bar{\zeta}^{b}\biggr|_{r=r_{+}}^{r=\infty}=\frac{2\ell q\eta}{\ell_{\star}r_{+}\Sigma(x_{1},r_{+})}(1-\bar{a}^{2})\;,\label{eq:potentialqbtz}\eeq
for gauge potential (\ref{eq:Agaugbarred}) and horizon generator  (\ref{eq:cannormgenerator}).\footnote{In the limit $\ell\to\infty$ the potential (\ref{eq:potentialqbtz}) reduces to the electrostatic potential for Kerr-Newman-AdS$_{4}$, e.g., Eq. (20) of \cite{Caldarelli:1999xj}, where in addition to the rescalings in Footnote \ref{fn:tenlimit}, also send $q\to \hat{q}/(\ell x_{1}^{2})$. Likewise, using $g_{\star}=\frac{\sqrt{16\pi G_{4}}}{\ell_{\star}}$, the charge (\ref{eq:elecchargeqBTZ}) becomes $Q\to \frac{Q\ell_{\star}}{2G_{4}\Xi}$ for $\Xi\equiv \left(1-\frac{\hat{a}^{2}}{\ell_{4}^{2}}\right)$, the physical charge of the Kerr-Newman-AdS$_{4}$ black hole (see Eq. (19) of \cite{Caldarelli:1999xj}, whereupon one further sets $G_{4}=1$ and $\ell_{\star}=2$).} Note that although we used the bulk gauge potential, the $x$-dependence drops out.

Since the original AdS$_{4}$ metric is an exact solution to Einstein-Maxwell theory, we emphasize that the brane metric (\ref{eq:qbtzmet}) is an exact solution to the semi-classical induced theory from the brane perspective. That is, the metric (\ref{eq:qbtzmet}) is an \emph{exact} quantum black hole in AdS$_{3}$, namely, the charged and rotating quantum BTZ solution. The title, in part, comes from the solution's relation studying perturbative backreaction in semi-classical Einstein-Hilbert-AdS$_{3}$ gravity, where the leading order contribution to the renormalized quantum stress-tensor $\langle T_{ij}\rangle$ has precisely the same form as in the perturbative computation (see Appendix \ref{app:scgravrev}). Notably, the strength of backreaction is controlled by the parameter $\ell\gg L_{\text{Planck}}$, such that $\ell\to0$ corresponds to the limit of vanishing backreaction due to the CFT. Further, with the exact solution one can compute the stress-tensor, in principle, to all orders in $\ell$, a distinct advantage compared to the non-holographic computation.

Using the expansion (\ref{eq:Tij0gen}) and (\ref{eq:Tij2gen}), let us compute the components of the quantum stress-tensor up to order $\ell^{2}$. As with the neutral rotating qBTZ black hole, the strategy is to first work in naive coordinates (\ref{eq:naivemetrot}), and then transform to canonically normalized coordinates. To this end, the components of the stress tensor at leading order (\ref{eq:Tij0gen}) are
\begin{align}
    \langle T\indices{^t_t}\rangle_0&=  \langle T\indices{^r_r}\rangle_0=\frac{2\mu \ell r+q^2\ell^2}{32\pi G_3 r^4}\;,
    \\
\langle T\indices{^\phi_\phi}\rangle_0&=\frac{-4\mu \ell r+7q^2 \ell^2}{32\pi G_3 r^4}\;,\\
\langle T\indices{^t_\phi}\rangle_0&=\frac{3\mu a \ell r-3a q^2 \ell^2}{16\pi G_3 r^6}\;.
\end{align}
Transforming to canonically normalized coordinates gives 
\begin{align}\notag
      \langle T\indices{^{\bar{t}}_{\bar {t}}}\rangle_0&=\frac{\ell}{16\pi G_3(1-\bar a^2)r^3} \left[\mu\left(1+2\bar a^2+\frac{3 \bar a^2 \ell_3^2}{x_1^2 r^2}\right)-\frac{q^2\ell}{r}\left(\frac{1-7 \bar a^2}{2}+\frac{3\bar a^2 \ell_3^2 }{x_1^2r^2}\right)\right]\;,\\ \notag
        \langle T\indices{^{\bar{r}}_{\bar{r}}}\rangle_0&=\frac{2\mu \ell r+q^2 \ell^2}{32\pi G_3 r^4}\;,\\
          \langle T\indices{^{\bar{\phi}}_{\bar{\phi}}}\rangle_0&=-\frac{\ell}{16\pi G_3(1-\bar a^2)r^3}\left[\mu\left(2+\bar a^2+\frac{3\bar a^2 \ell_3^2}{x_1^2 r^2}\right)-\frac{q^2 \ell}{r}\left(\frac{-7+ \bar a^2}{2}+\frac{3\bar a^2 \ell_3^2 }{x_1^2 r^2}\right)\right]\;,\\ \notag
            \langle T\indices{^{\bar{t}}_{\bar{\phi}}}\rangle_0&=-\frac{3 \ell \bar a \ell_3}{16\pi G_3(1-\bar a^2)r^3}\left(\mu-\frac{q^2\ell}{r}\right)\left(1+\frac{\bar a^2 \ell_3^2}{x_1^2 r^2}\right)\;,\\ \notag
            \langle T\indices{^{\bar{\phi}}_{\bar{t}}}\rangle_0&=\frac{ 3\ell \bar a}{16\pi G_3(1-\bar a^2)\ell_3 r^3}\left(\mu-\frac{q^2\ell}{r}\right)\left(1+\frac{\ell_3^2}{x_1^2 r^2}\right)\;.
\end{align}
In the vanishing charge and momentum limit, we recover the quantum stress-tensor supporting the neutral, static qBTZ black hole (upon where $r=\eta^{-1}\bar{r}$), while with rotation we recover the quantum stress-tensor supporting the neutral rotating qBTZ black hole \cite{Emparan:2020znc}.

For completeness, the components of the stress-tensor at order $\ell^{2}$ given in (\ref{eq:Tij2gen}), in naive coordinates, are
\begin{align}\notag
    8\pi G_3\langle T\indices{^{t}_{t}}\rangle_2&=
    \begin{aligned}[t]
    &\frac{\ell(21\ell \ell_3^2q^2(16a^2+5q^2 \ell^2)-\ell_3^2 \mu (90a^2+121q^2 \ell^2))}{32\ell_3^2 r^8}\\
     &+\frac{\ell(\ell \ell_3^2(19 \mu^2+96k q^2)r^2-18k \mu \ell_3^2 r^3+70\ell q^2 r^4-11\mu r^5)}{32\ell_3^2 r^8}\;,
    \end{aligned} \\\notag
 8\pi G_3  \langle T\indices{^{r}_{r}}\rangle_2&= 
   \begin{aligned}[t]
    &\frac{a\ell(33\ell^3 \ell_3^2 q^4-37\ell^2 \ell_3^2 \mu q^2 r+\ell \ell_3^2(7\mu^2+24k q^2)r^2)}{32\ell_3^2 r^8}\\
     &+\frac{\ell(-2\ell q^2 r^4+6a^2 \ell_3^2(16q^2 \ell-5\mu r)+\mu r^3(-6k \ell_3^2+r^2))}{32\ell_3^2 r^8}\;,
    \end{aligned}\\
    8\pi G_3 \langle T\indices{^{\phi}_{\phi}}\rangle_2&=  \begin{aligned}[t]
    &\frac{\ell(-3\ell \ell_3^2q^2(144a^2+53q^2 \ell^2)+6\ell_3^2 \mu (20a^2+29q^2 \ell^2))}{32\ell_3^2 r^8}\\
     &+\frac{\ell(-\ell \ell_3^2(29 \mu^2+120 k q^2)r^2+24 k \mu \ell_3^2 r^3-66\ell q^2 r^4+10\mu r^5)}{32\ell_3^2 r^8}\;,
    \end{aligned}\\\notag
   8\pi G_3 \langle T\indices{^{t}_{\phi}}\rangle_2&=\frac{6a\ell(-7 \ell q^2+2\mu r)}{r^6}\;,\\\notag
   8\pi G_3 \langle T\indices{^{\phi}_{t}}\rangle_2&=  \begin{aligned}[t]
    &\frac{a\ell(48\ell \ell_3^2 q^2(16a^2 +9q^2 \ell^2)-7\mu \ell_3^2(30a^2+73q^2 \ell^2)r)+}{32\ell_3^2 r^8}\;,\\
     &\frac{\ell(96\ell \ell_3^2(\mu^2+ 4kq^2)r^2-90k \mu \ell_3^2r^3+304\ell q^2 r^4-69\mu r^5)}{32\ell_3^2 r^8}\;.
    \end{aligned}
\end{align}
One can rewrite these components in terms of canonically normalized coordinates, however, the resulting expressions are cumbersome and unrevealing.

\vspace{2mm}

\noindent \textbf{Quantum black hole mechanics and entropy.} With the horizon angular velocity (\ref{eq:horangvel}), surface gravity (\ref{eq:surfacegrav}), mass (\ref{eq:massqbtz}), rotation (\ref{eq:angmomeqbtz}), electric charge (\ref{eq:elecchargeqBTZ}), and potential (\ref{eq:potentialqbtz}), we have nearly every quantity characterizing the mechanics of the quantum BTZ black hole. All that remains is the Bekenstein-Hawking area formula. To this end, we compute the Bekenstein-Hawking area formula for the AdS$_{4}$ C-metric, however, working in barred coordinates $(\bar{t},r,\bar{\phi})$, such that we leave the radial coordinate untouched. 
Consequently, the line element for the charged and rotating AdS$_{4}$ C-metric becomes
\beq
\begin{split} 
ds^{2}&=\frac{\ell^{2}}{(\ell+xr)^{2}}\biggr\{-\frac{H(r)}{\Sigma(x,r)}\left[\eta(d\bar{t}-\bar{a}\ell_{3}d\bar{\phi})+ax^{2}\eta\left(d\bar{\phi}-\frac{\bar{a}}{\ell_{3}}d\bar{t}\right)\right]^{2}+\frac{\Sigma(x,r)}{H(r)}dr^{2}\\
&+r^{2}\frac{\Sigma(x,r)}{G(x)}dx^{2}+\frac{r^{2}G(x)}{\Sigma(x,r)}\left[\eta\left(d\bar{\phi}-\frac{\bar{a}}{\ell_{3}}d\bar{t}\right)-\frac{a}{r^{2}}\eta(d\bar{t}-\bar{a}\ell_{3}d\bar{\phi})\right]^{2}\biggr\}\;.
\end{split}
\eeq
At a fixed $r-\bar{t}$ surface (with $r=r_{+}$), the line element of the induced metric at the codimension-2 surface is 
\beq 
\begin{split}
 ds^{2}_{(2)}=\frac{\ell^{2}}{(\ell+xr)^{2}}\left[r_{+}^{2}\frac{\Sigma(x,r_{+})}{G(x)}dx^{2}+r_{+}^{2}\frac{G(x)}{\Sigma(x,r_{+})}\eta^{2}d\bar{\phi}^{2}\left(1+\frac{a^{2}x_{1}^{2}}{r_{+}^{2}}\right)^{2}\right]\;,
\end{split}
 \eeq
 where we used $H(r_{+})=0$ and $\bar{a}=ax_{1}^{2}/\ell_{3}$. The codimension-2 area element is 
 \beq \sqrt{\gamma}d^{2}x=\frac{\ell^{2}}{(\ell+xr)^{2}}r_{+}^{2}\eta \left(1+\frac{a^{2}x_{1}^{2}}{r_{+}^{2}}\right)dxd\bar{\phi}\;.\eeq
 Hence, the bulk Bekenstein-Hawking area formula is 
 \beq 
 \begin{split} 
 \frac{\text{Area}^{(4)}(r_{+})}{4G_{4}}&=\frac{2}{4G_{4}}\int_{0}^{2\pi}d\phi\int_{0}^{x_{1}}dx\frac{r_{+}^{2}\ell^{2}}{(\ell+xr_{+})^{2}}\eta\left(1+\frac{a^{2}x_{1}^{2}}{r_{+}^{2}}\right)\\
 &=\frac{\pi}{G_{4}}\eta\ell x_{1}\frac{(r_{+}^{2}+a^{2}x_{1}^{2})}{\ell+r_{+}x_{1}}\;,
 \end{split}
 \label{eq:bulkent}\eeq
 where the factor of two in the first equality is due to the brane being two-sided. 
 
With the area-formula, it can be verified that the black hole quantities obey the following first law of mechanics
\beq \delta M=\frac{\kappa_{+}}{2\pi G_{4}}\delta\left(\frac{\text{Area}^{(4)}(r_{+})}{4G_{4}}\right)+\bar{\Omega}_{+}\delta J+\Phi\delta Q\;.\label{eq:bhmechfirstlaw}\eeq
This is unsurprising from the bulk perspective as this is nothing more than the first law of mechanics for the regulated AdS$_{4}$ black hole. As it stands, the first law (\ref{eq:bhmechfirstlaw}) is suggestive of a thermodynamic interpretation, where the the area formula (\ref{eq:bulkent}) is identified with entropy $S_{\text{BH}}^{(4)}$. We will justify this viewpoint in the next section by explicitly deriving each of these quantities from the gravitational partition function. 

Before moving on, let us briefly comment on the interpretation of each quantity from the brane perspective. As noted already, all quantities here are identified with those of the quantum black hole. Notably, from the brane point of view, the classical entropy $S_{\text{BH}}^{(4)}$  must be a sum of gravitational entropy and the von Neumann entropy of the holographic CFT$_{3}$. In other words,  the bulk entropy is identified with the semi-classical \emph{generalized} entropy in three-dimensions
\beq S_{\text{BH}}^{(4)}\equiv S_{\text{gen}}^{(3)}\;.\label{eq:S4Sgenqbtz}\eeq
Since the brane gravity is in general a higher-derivative theory, the gravitational contribution to the generalized entropy is captured by the Iyer-Wald entropy functional \cite{Wald:1993nt,Iyer:1994ys},
\beq S_{\text{IW}}\equiv-2\pi\int_{\mathcal{H}}\hspace{-1mm} d^{d-2}x\sqrt{q}\;\frac{\partial\mathcal{L}}{\partial R^{ijkl}}\epsilon_{ij}\epsilon_{kl}\;,\eeq
where $q_{ij}$ is the induced metric of the codimension-2 cross-section of the horizon $\mathcal{H}$ with binormal $\epsilon_{ij}$, and $\mathcal{L}$ is the Lagrangian. Since the leading order contribution to the induced brane action is simply the Einstein-Hilbert-AdS$_{3}$ term, the leading contribution to the gravitational entropy is the 3D Bekenstein-Hawking entropy formula, $S^{(3)}_{\text{BH}}$. The next contribution to the Iyer-Wald entropy enters at order $\ell^{2}$, while the CFT von Neumann entropy enters at order $\ell$, as schematically written in (\ref{eq:entintro}). We emphasize that the  CFT$_{3}$ von Neumann entropy can be explicitly found by taking the difference $S_{\text{CFT}}=S^{(3)}_{\text{gen}}-S_{\text{IW}}^{(3)}$, in principle to all orders in $\ell^{2}$ since the identification (\ref{eq:S4Sgenqbtz}) is exact.


\section{Euclidean action of quantum BTZ black holes}\label{sec:qBHthermo}

In the previous section we saw how the mechanical quantities of the bulk AdS$_{4}$ black hole obey a mechanical first law. Further, via double holography, such quantities have a dual interpretation from the brane perspective. This observation prompts a thermodynamic interpretation, where, notably, the classical area-entropy of the bulk black hole is identified with the generalized entropy of the quantum BTZ black hole. An explicit derivation of the thermodynamic quantities beyond the neutral and static quantum BTZ, however, in thus far lacking in the literature. Here, we fill this gap. We will adapt the Gibbons-Hawking prescription \cite{Gibbons:1976ue} and explicitly derive the thermodynamic variables from the grand canonical gravitational partition function of the AdS$_{4}$  C-metric endowed with a Karch-Randall ETW brane. 

\subsection{General strategy}

A first principles approach to gravitational thermodynamics is via the path integral formulation of Euclidean quantum gravity. In this context, one follows the Gibbons and Hawking prescription for constructing the canonical gravitational partition function \cite{Gibbons:1976ue},
\beq \mathcal{Z}(\beta)=\int Dg D\varphi e^{-I_{E}[g,\varphi]}\;.\label{eq:gravpartgen}\eeq
Here $\beta$ denotes the (inverse) temperature, held fixed to define the canonical ensemble, and 
the integration measure indicates a sum over all dynamical (Euclidean) metrics and matter field configurations $\varphi$ in a theory characterized by its Euclidean action, $I_{E}[g,\varphi]$.
In practice, the partition function is evaluated perturbatively by expanding $I_{E}$ about solutions $\{g_{c},\varphi_{c}\}$ to the classical equations of motion. For each saddle-point, the action expands as $I_{E}[g,\varphi]=I^{\text{on-shell}}_{E}[g_{c},\varphi_{c}]+I^{(2)}_{E}[h,\psi]+...$, for small fluctuations $\{h,\psi\}$ about the classical saddles, and 
 $I^{(2)}_{E}[h,\psi]$ is quadratic in field fluctuations giving the leading (1-loop) perturbative quantum correction to $I_{E}[g,\varphi]$.  Heuristically, the partition function (\ref{eq:gravpartgen}) expands as
\beq
\begin{split}
\mathcal{Z}(\beta)&\approx \sum_{\{g_{c},\varphi_{c}\}} e^{-I_{E}^{\text{on-shell}}[g_{c},\varphi_{c}]}\int Dh D\psi e^{-I_{E}^{(2)}[h,\psi]+...}\equiv\sum_{\{g_{c},\varphi_{c}\}}Z^{(0)}[g_{c},\varphi_{c}]Z^{(1)}[g_{c},\varphi_{c}]+...\;\;.
\end{split}
\eeq
In this article we will only focus on the thermodynamic quantities coming from the leading tree-level partition function $Z^{(0)}$, 
\beq \mathcal{Z}(\beta)\approx Z^{(0)}(\beta)=e^{-I^{\text{on-shell}}_{E}}\;.\label{eq:partfuncgen}\eeq
It would be interesting to uncover the leading perturbative quantum corrections by evaluating the 
the 1-loop partition function, $Z^{(1)}$, however, we will not pursue this here. 

Underlying the set-up is the assumption our thermal system is near equilibrium. As with de Sitter black holes, which generally have at least two horizons of differing temperature, thermodynamics of the C-metric is generally subtle due to the presence of both acceleration and black hole horizons. It is not obvious how an accelerating black hole can be in thermal equilibrium. The advantage of working with the AdS$_{4}$ C-metric is that it describes a slowly accelerating black hole \cite{Podolsky:2002nk}, cf. (\ref{eq:restricads4l}). This will allow us to unambiguously define a system in equilibrium. Further, due to the brane construction, we have removed the influence of the cosmic string which would otherwise need to be accounted for \cite{Appels:2016uha,Appels:2017xoe,Anabalon:2018ydc,Appels:2018jcs}. This makes our thermodynamic analysis less subtle than the standard C-metric.

Our goal then is to evaluate the bulk on-shell Euclidean action of our theory. So far this includes the Euclidean Einstein-Maxwell-$\Lambda$ action, a Gibbons-Hawking-York (GHY) boundary term for the theory to have a well-posed variational problem in the presence of a boundary, and the brane action. Altogether, 
\beq 
\begin{split} 
I_{E}&=I_{\text{EM}}+I_{\text{GHY}}+I_{\mathcal{B}}\\
&=-\frac{1}{16\pi G_{4}}\int_{\mathcal{M}_{E}}\hspace{-4mm} d^{4}x\sqrt{\hat{g}}\left(\hat{R}+\frac{6}{\ell_{4}^{2}}-\frac{\ell_{\star}^{2}}{4}\hat{F}^{2}\right)-\frac{1}{8\pi G_{4}}\int_{\partial\mathcal{M}_{E}}\hspace{-4mm} d^{3}x\sqrt{h}K+\tau\int_{\mathcal{B}}d^{3}x\sqrt{h}\;,
\end{split}
\label{eq:Eucactpre}\eeq
where $\mathcal{M}_{E}$ denotes the Euclidean section of the four-dimensional background with asymptotic boundary $\partial\mathcal{M}_{E}$. We may append the above action by including an appropriate boundary term associated with the bulk Maxwell contribution in the event we choose to work in a fixed (electric) charge or `canonical' ensemble. Specifically, we would add 
\beq I_{A}=\pm \frac{\ell_{\star}^{2}}{64\pi G_{4}}\int_{\partial\mathcal{M}_{E}} \hspace{-4mm} d^{3}x\sqrt{h}n_{a}\hat{F}^{ab}A_{b}\;,\eeq
where $\pm$ depends on our choice of inward or outward pointing normal vector $n_{a}$. This term is needed assuming the (non-Dirichlet) boundary conditions, $\delta (\sqrt{h}n_{a}\hat{F}^{ab}A_{b})_{\partial\mathcal{M}}=0$ \cite{Hawking:1995ap}.

As it stands, the action above will have an infrared divergence when evaluated on-shell. Two common ways  to regulate this divergence include background subtraction, or introduce a local counterterm action that respects the boundary conditions.  Following \cite{Kudoh:2004ub,Panella:2024sor}, here we take a different approach. Rather, we will add a second brane action, associated with an ETW brane that we place at the second distinct umbilic surface of the C-metric (see below), 
and a GHY boundary term associated with the brane at $x=0$. In other words, in addition to the bulk Einstein-Maxwell action $I_{\text{EM}}$, we will also have a brane and GHY action for each umbilic hypersurface.

To this end, it proves useful to rewrite the C-metric (\ref{eq:rotatingCmetqbtz}) in coordinates $(t,y,x,\phi)$ where we perform the transformation
\beq t\to t\ell\;,\quad r\to -\frac{\ell}{y}\;,\label{eq:coordscalings}\eeq
such that $t$ and $y$ are dimensionless, and make the identifications
\beq \lambda\equiv \frac{\ell^{2}}{\ell_{4}^{2}}-1=\frac{\ell^{2}}{\ell_{3}^{2}}\;,\quad k=-\kappa\;, \quad a\to a\ell \;,\label{eq:paramscals}\eeq
for dimensionless $a$.
In these coordinates, it is evident the hypersurfaces at $x=0$ and $y=0$ are totally umbilic (see Appendix \ref{app:cmetgeom} for details).  
We imagine putting a purely tensional brane with action (\ref{eq:genbraneact}) at each hypersurface, such that imposing Israel's junction conditions (\ref{eq:israeljuncconds}) gives the tensions
\beq \tau_{x}=\frac{1}{2\pi G_{4}\ell}\;,\quad \tau_{y}=\frac{\sqrt{\lambda}}{2\pi G_{4}\ell}\;.\label{eq:branetensionsxy}\eeq

The total Euclidean action $I_{E}$ characterizing the bulk Riemannian spacetime $\mathcal{M}_{E}$ endowed with Euclidean metric $\hat{g}$, and branes $\mathcal{B}_{x}$ and $\mathcal{B}_{x}$ at $x,y=0$, is then given by 
\beq I=I_{\text{EM}}+I_{\text{GHY}}^{(x)}+I_{\text{GHY}}^{(y)}+I_{\mathcal{B}_{x}}+I_{\mathcal{B}_{y}}\;.\label{eq:Eucactgen}\eeq
The Einstein-Maxwell bulk action is (\ref{eq:Eucactpre}), while the two GHY boundary terms are
\beq I_{\text{GHY}}^{(x)}=-\frac{1}{8\pi G_{4}}\int_{\mathcal{B}_{x}}d^{3}x\sqrt{h_{(x)}}K^{(x)}\;,\qquad I_{\text{GHY}}^{(y)}=-\frac{1}{8\pi G_{4}}\int_{\mathcal{B}_{y}}d^{3}x\sqrt{h_{(y)}}K^{(y)}\;,\label{eq:GHYtermsapp}\eeq
and the brane actions have the form
\beq I_{\mathcal{B}_{x}}=-\tau_{x}\int_{\mathcal{B}_{x}}d^{3}x\sqrt{h_{(x)}}\;,\qquad I_{\mathcal{B}_{y}}=-\tau_{y}\int_{\mathcal{B}_{y}}d^{3}x\sqrt{h_{(y)}}\;.\label{eq:BxByact}\eeq
Here $h_{(x)}$ and $K^{(x)}$ denote the induced metric and trace of the extrinsic curvature of the $x=0$ hypersurface, and similarly for $h_{(y)}$ and $K^{(y)}$.

\subsection{Regularity of the accelerating black hole and thermal ensembles} 

To use the framework of Euclidean gravity, we Wick rotate the Lorentzian time and angular and charge parameters such that the Euclideanized metric is Riemannian and the gauge field is real. In particular, under
\beq t\to -it_{E}\;,\quad a\to ia_{E}\;,\quad q\to iq_{E}\;,\label{eq:Wickrot}\eeq
for Euclidean time $t_{E}$ and $a_{E},q_{E}\in\mathbb{R}$, the AdS$_{4}$ C-metric in $(t,y,x,\phi)$ coordinates (see (\ref{eq:rotCmetapp})) becomes
\beq 
\begin{split}
ds^{2}_{E}&=\frac{\ell^2}{(x-y)^{2}}\biggr[\frac{H(y)}{\Sigma(x,y)}(dt_{E}-a_{E}x^{2}d\phi)^{2}+\frac{\Sigma(x,y)}{H(y)}dy^{2}+\frac{\Sigma(x,y)}{G(x)}dx^{2}+\frac{G(x)}{\Sigma(x,y)}(d\phi-a_{E}y^{2}dt_{E})^{2}\biggr]\;,
\end{split}
\label{eq:rotCmetEuc}\eeq
with metric functions
\beq
\begin{split}
&H(y)=\lambda-ky^{2}+\mu y^{3}-a_{E}^{2}y^{4}-q_{E}^{2}y^{4}\;,\quad \Sigma(x,y)=1-a_{E}^{2}x^{2}y^{2}\\
&G(x)=1+kx^{2}-\mu x^{3}-a_{E}^{2}\lambda x^{4}+q_{E}^{2}x^{4}\;.
\end{split}
\eeq
and real gauge field 
\beq A_{E}=A_{a}dx^{a}=\frac{2\ell q_{E} y}{\ell_{\star}(1-a_{E}^{2}x^{2}y^{2})}(dt_{E}-a_{E}x^{2}d\phi)\;.\label{eq:gaugefieldEuc}\eeq
As it stands, in the current choice of gauge, $A_{a}A^{a}$ diverges on the horizon. We will return to this point momentarily.

The Riemannian geometry (\ref{eq:rotCmetEuc}) has two types of conical singularities. First, there are the familiar ones arising due to the zeros of $G(x)$. As before, we eliminate the zero at $x=x_{1}$ via the identification $\phi\sim \phi+\Delta\phi$ with $\Delta\phi=4\pi/|G'(x_{1})|$ such that,
\beq (\bar{t}_{E},\bar{\phi})\sim (\bar{t}_{E},\bar{\phi}+2\pi)\;.\label{eq:cannorid1}\eeq
Here, it is useful to know the canonically normalized coordinates $(\bar{t},\bar{\phi})$ under the continuation (\ref{eq:Wickrot}) rotate to
\beq 
\begin{split} 
&\bar{t}\to -i\bar{t}_{E}\;,\quad \bar{t}_{E}\equiv \frac{1}{(1+\bar{a}_{E}^{2}\ell^{2})\eta}t_{E}-\frac{\bar{a}_{E}\ell_{3}}{(1+\bar{a}_{E}^{2}\ell^{2})\eta}\phi\;,\\
&\bar{\phi}\to \frac{\bar{a}_{E}\ell^{2}}{\ell_{3}(1+\bar{a}_{E}^{2}\ell^{2})\eta}t_{E}+\frac{1}{(1+\bar{a}_{E}^{2}\ell^{2})\eta}\phi\;,
\end{split}
\label{eq:barrcoordEuc}\eeq
from which the identification (\ref{eq:cannorid1}) follows via $t_{E}\sim t_{E}+2\pi \bar{a}_{E}\ell_{3}\eta$ and $\phi\sim \phi+2\pi\eta$.

The second conical singularity is located at the black hole horizon $y=y_{+}$, where $H(y_{+})=0$. To see this, zoom in to the near-horizon region, where the Euclidean black hole (\ref{eq:rotCmetEuc}) has the line element
\beq
\begin{split} 
ds^{2}_{E}&\approx \frac{\ell^{2}}{(x-y_{+})^{2}}\biggr[\Sigma(x,y_{+})\left(d\rho^{2}+\frac{(H'(y_{+}))^{2}\rho^{2}}{4\Sigma(x,y_{+})^{2}}[dt_{E}-a_{E}x^{2}d\phi]^{2}\right)\\
&+\frac{\Sigma(x,y_{+})}{G(x)}dx^{2}+\frac{G(x)}{\Sigma(x,y_{+})}(d\phi-a_{E}y_{+}^{2}dt_{E})^{2}\biggr]\;,
\end{split}
\label{eq:nearhorqbh}\eeq
where we expanded $H(y)\approx H'(y_{+})(y-y_{+})+\mathcal{O}((y-y_{+})^{2})$ and introduced the radial coordinate $(y-y_{+})\equiv \frac{H'(y_{+})\rho^{2}}{4}$. To ensure regularity by removing the conical singularity at the horizon $\rho=0$, we find the Euclidean time $t_{E}$ and angular variable $\phi$ must have periodicities
\beq
\begin{split} 
&t_{E}\sim t_{E}+\Delta t_{E}\;,\quad \Delta t_{E}=\frac{4\pi}{H'(y_{+})}\;,\\
&\phi\sim \phi+\delta \phi\;,\quad \delta \phi =a_{E}y_{+}^{2}\Delta t_{E}\;.
\label{eq:periodrotmain}
\end{split}\eeq
For more details, see Appendix \ref{app:cmetgeom}. Combining the coordinates (\ref{eq:barrcoordEuc}) with the periodicity (\ref{eq:periodrotmain}) yields
\beq
\begin{split} 
&\bar{t}_{E}\sim 
\bar{t}_{E}+\frac{1}{(1+\bar{a}_{E}^{2}\ell^{2})\eta}\frac{4\pi\Sigma(x_{1},y_{+})}{H'(y_{+})}\;,\\
&\bar{\phi}\sim
\bar{\phi}+\frac{1}{(1+\bar{a}_{E}^{2}\ell^{2})\eta}\left(\frac{a_{E}x_{1}^{2}\ell^{2}}{\ell_{3}^{2}}+a_{E}y_{+}^{2}\right)\frac{4\pi}{H'(y_{+})}\;.
\end{split}
\label{eq:rotvariablesEuccannorm}\eeq
In the static limit ($a_{E}=0$), we recover the periodicities in \cite{Kudoh:2004ub,Panella:2024sor}. Further, notice that all of the dependence on the charge parameter $q_{E}$ is implicitly in $H'(y_{+})$ and $\eta$. 

\vspace{2mm}

\noindent \textbf{Thermal ensembles.} Ordinarily, thermal quantum field theory at finite temperature $T\equiv \beta^{-1}$ is equivalent to Euclidean QFT with a time periodicity $t_{E}\sim t_{E}+\beta$. This follows at the level of thermal correlation functions, where, for example, Euclidean Green functions are periodic (for bosons) or anti-periodic (for fermions) in Euclidean time $t_{E}$ with period $\beta$, the so-called Kubo-Martin-Schwinger (KMS) condition. This leads to one expressing the canonical thermal partition function $\mathcal{Z}(\beta)$ as a Euclidean path integral, where the Euclidean time is compactified to a circle $S^{1}_{\beta}$. Fixing the thermodynamic data defining the ensemble corresponds to fixing the boundary conditions of the fields in the path integral. 

In a grand canonical ensemble, where chemical potentials of conserved charges are specified in addition to temperature, the analog of the KMS condition implies a twisted identification. In particular,  for an angular momentum $J$ generating rotations along an angle $\bar{\phi}$, one has (in canonically normalized coordinates)
\beq (\bar{t}_{E},\bar{\phi})\sim (\bar{t}_{E}+\beta,\bar{\phi}-i\beta\Omega)\;,\label{eq:KMSid}\eeq
for inverse temperature $\beta$ and potential $\Omega$. In this ensemble, both the temperature and potential are held fixed, such that the grand canonical partition function is a function of the thermodynamic data, $\mathcal{Z}(\beta,\Omega)$.

Applying these tenets of thermal field theory to the gravitational field, combining the identification to make the Euclidean geometry regular at the horizon (\ref{eq:rotvariablesEuccannorm}) together with (\ref{eq:KMSid}), we identify the inverse temperature and potential to be\footnote{Since we work with dimensionless time $\bar{t}_{E}$, the twisted KMS condition we use here is $(\bar{t}_{E},\bar{\phi})\sim(\bar{t}_{E}+\beta\ell^{-1},\bar{\phi}-i\beta\Omega)$.}
\beq \beta\equiv\frac{\ell\Sigma(x_{1},y_{+})}{(1+\bar{a}_{E}^{2}\ell^{2})\eta}\frac{4\pi}{H'(y_{+})}\;.\label{eq:invtemp}\eeq
\beq i\Omega=-\frac{a_{E}}{\ell\Sigma(x_{1},y_{+})}\left(\frac{x_{1}^{2}\ell^{2}}{\ell_{3}^{2}}+y_{+}^{2}\right)\;.\label{eq:Omeucc}\eeq
Undoing the Wick rotation ($a_{E}=-ia$) and using $y_{+}=-\ell/r_{+}$ with $a \to a/\ell$, we find
\beq \beta=\frac{\Sigma(x_{1},r_{+})}{(1-\bar{a}^{2})\eta}\frac{4\pi}{H'(r_{+})}=\frac{2\pi}{\bar{\kappa}_{+}}\;,\eeq
for surface gravity $\bar{\kappa}_{+}$ (\ref{eq:surfacegrav}). Thus, $\beta$ is equal to the inverse Hawking temperature. Further, we find $\Omega=\bar{\Omega}_{+}$ for horizon angular velocity $ \bar{\Omega}_{+}$ (\ref{eq:horangvel}). 

 Lastly, we have a conserved electric charge due to a $U(1)$ gauge field $A$ and corresponding chemical potential $\Phi$. In the grand canonical ensemble, the potential $\Phi$ is taken to be fixed, while in the canonical ensemble one instead fixes the charge. The potential is also found by imposing regularity of the gauge field at the horizon. Moving to canonically normalized coordinates, the Euclideanized gauge field is
 \beq \bar{A}_{E}=\frac{2\ell q_{E}y\eta}{\ell_{\star}(1-a^{2}_{E}x^{2}y^{2})}\left[d\bar{t}_{E}\left(1+\frac{a_{E}x^{2}\bar{a}_{E}\ell^{2}}{\ell_{3}}\right)+d\bar{\phi}\left(\bar{a}_{E}\ell_{3}-a_{E}x^{2}\right)\right]\;.\label{eq:gaugpotbarr}\eeq
 To ensure regularity at the horizon, we shift $\bar{A}_{E}\to \bar{A}_{E}+C_{E}d\bar{t}_{E}$, where $C_{E}=-A_{\bar{t}_{E}}|_{y=y_{+}}$, from which we identify the (Euclidean) electric potential $C_{E}=\ell\Phi_{E}$ with 
 \beq \Phi_{E}=-\frac{2q_{E}y_{+}\eta}{\ell_{\star}(1-a_{E}^{2}x_{1}^{2}y_{+}^{2})}\left(1+\bar{a}_{E}^{2}\ell^{2}\right)\;.\eeq
 Undoing the Wick rotation (\ref{eq:Wickrot}) and coordinate and parameter rescalings (\ref{eq:coordscalings}) and (\ref{eq:paramscals}), we recover $\Phi_{E}=i\Phi$ with $\Phi$ given in (\ref{eq:potentialqbtz}).

\subsection{Euclidean action}

We now turn to the evaluation of the action (\ref{eq:Eucactgen}) of the Euclidean black hole (\ref{eq:rotCmetEuc}). Specifically, we will work in the grand canonical ensemble, i.e., an ensemble of fixed temperature $\beta^{-1}$ and potentials $\Omega$ and $\Phi$. 

\vspace{2mm}

\noindent \textbf{Bulk action.} We begin with the Einstein-Hilbert-Maxwell bulk contribution, $I_{\text{EM}}=I_{\text{EH}}+I_{\text{Max}}$. Explicitly, 
\beq 
\begin{split} 
I_{\text{EH}}&=-\frac{1}{16\pi G_{4}}\int_{\mathcal{M}_{E}}\hspace{-4mm} d^{4}x\sqrt{\hat{g}}\left(\hat{R}+\frac{6}{\ell_{4}^{2}}\right)=\frac{6\ell^{4}}{16\pi G_{4}\ell_{4}^{2}}\int_{0}^{\Delta t_{E}}\hspace{-2mm} dt_{E}\int_{0}^{\Delta\phi}\hspace{-2mm} d\phi \int_{\epsilon_{x}}^{x_{1}}\hspace{-2mm} dx\int_{y_{+}}^{\epsilon_{y}}\hspace{-2mm} dy\frac{(1-a_{E}^{2}x^{2}y^{2})}{(x-y)^{4}}\\
&=\frac{\ell^{4}}{16\pi G_{4}\ell_{4}^{2}}\Delta t_{E}\Delta\phi\left[\frac{1}{(x_{1}-y_{+})^{2}}-\frac{1}{(y_{+}-\epsilon_{x})^{2}}-\frac{1}{(x_{1}-\epsilon_{y})^{2}}+\frac{1}{(\epsilon_{x}-\epsilon_{y})^{2}}\right]\\
&+\frac{a_{E}^{2}\ell^{4}}{16\pi G_{4}\ell_{4}^{2}}\Delta t_{E}\Delta\phi(x_{1}-\epsilon_{x})\left[\frac{\epsilon_{y}^{3}(\epsilon_{y}(x_{1}+\epsilon_{x})-2x_{1}\epsilon_{x})}{(x_{1}-\epsilon_{y})^{2}(\epsilon_{x}-\epsilon_{y})^{2}}-\frac{y_{+}^{3}(x_{1}(y_{+}-2\epsilon_{x})+y_{+}\epsilon_{x})}{(x_{1}-y_{+})^{2}(y_{+}-\epsilon_{x})^{2}}\right]\;.
\end{split}
\eeq
and
\beq
\begin{split} 
I_{\text{Max}}&=\frac{\ell_{\star}^{2}}{64\pi G_{4}}\int_{\mathcal{M}_{E}}\hspace{-4mm} d^{4}x\sqrt{\hat{g}}\hat{F}^{2}=\frac{q_{E}^{2}\ell^{2}}{8\pi G_{4}}\Delta t_{E}\Delta \phi\int_{\epsilon_{x}}^{x_{1}}dx\int_{y_{+}}^{\epsilon_{y}}dy\frac{(1+6a_{E}^{2}x^{2}y^{2}+a_{E}^{4}x^{4}y^{4})}{(1-a_{E}^{2}x^{2}y^{2})^{3}}\\
&=\frac{q_{E}^{2}\ell^{2}}{8\pi G_{4}}\Delta t_{E}\Delta\phi\left[\frac{y_{+}\epsilon_{x}}{1-a_{E}^{2}y_{+}^{2}\epsilon_{x}^{2}}+\frac{x_{1}\epsilon_{y}}{1-a_{E}^{2}x_{1}^{2}\epsilon_{y}^{2}}-\frac{x_{1}y_{+}}{1-a_{E}^{2}x_{1}^{2}y_{+}^{2}}-\frac{\epsilon_{x}\epsilon_{y}}{1-a_{E}^{2}\epsilon_{x}^{2}\epsilon_{y}^{2}}\right]\;.
\end{split}
\eeq
Here we introduced hard cutoffs at $x=\epsilon_{x}\ll1$ and $y=\epsilon_{y}\ll1$, where at the end of the computation we will take the limit $\epsilon_{x,y}\to0^{+}$. In particular, notice that the Einstein-Hilbert term diverges as $\epsilon_{x,y}\to0$; specifically, the term independent of the angular rotation parameter $a_{E}$. Meanwhile, we observe the Maxwell term remains finite as $\epsilon_{x,y}\to0^{+}$.

\vspace{2mm}

\noindent \textbf{Boundary actions.} Let us now evaluate the GHY boundary and brane actions (\ref{eq:GHYtermsapp}) and (\ref{eq:BxByact}). First note that
\beq I^{(x)}_{\text{GHY}}=-\frac{3}{4}I_{\mathcal{B}_{x}}\;,\quad I^{(y)}_{\text{GHY}}=-\frac{3}{4}I_{\mathcal{B}_{y}}\;, \eeq
for $K^{(x)}=-\frac{3}{\ell}$, $K^{(y)}=-\frac{3\sqrt{\lambda}}{\ell}$ and brane tensions (\ref{eq:branetensionsxy}) and
\beq 
\begin{split}
I_{\mathcal{B}_{x}}&=-\frac{1}{2\pi G_{4}}\int_{\mathcal{B}_{x}}d^{3}x\sqrt{h_{(x)}}=-\frac{\ell^{2}}{2\pi G_{4}}\Delta t_{E}\Delta \phi\int_{y_{+}}^{\epsilon_{y}}dy\frac{1}{(\epsilon_{x}-y)^{3}}\\
&=-\frac{\ell^{2}}{4\pi G_{4}}\Delta t_{E}\Delta\phi\left[\frac{1}{(\epsilon_{x}-\epsilon_{y})^{2}}-\frac{1}{(\epsilon_{x}-y_{+})^{2}}\right]\;,
\end{split}
\eeq
\beq
\begin{split} 
I_{\mathcal{B}_{y}}&=-\frac{\sqrt{\lambda}}{2\pi G_{4}}\int_{\mathcal{B}_{y}}d^{3}x\sqrt{h_{(y)}}=\frac{\ell^{2}\lambda}{4\pi G_{4}}\Delta t_{E}\Delta\phi\left[\frac{1}{(x_{1}-\epsilon_{y})^{2}}-\frac{1}{(\epsilon_{x}-\epsilon_{y})^{2}}\right]\;,
\end{split}
\eeq
where we used $G(\epsilon_{x})\approx 1$, $H(\epsilon_{y})\approx \lambda$, and $\Sigma(\epsilon_{x},y),\Sigma(x,\epsilon_{y})\approx 1$. Combined, we have that the total boundary plus brane actions evaluate to 
\beq 
\begin{split} 
I_{\text{bdry}}&=\frac{1}{4}(I_{\mathcal{B}_{x}}+I_{\mathcal{B}_{y}})\\
&=-\frac{\ell^{2}}{16\pi G_{4}}\Delta t_{E}\Delta\phi\left[\frac{(1+\lambda)}{(\epsilon_{x}-\epsilon_{y})^{2}}-\frac{1}{(\epsilon_{x}-y_{+})^{2}}-\frac{\lambda}{(x_{1}-\epsilon_{y})^{2}}\right]\;.
\end{split}
\eeq

\vspace{2mm}

\noindent \textbf{Total on-shell action.} Implementing $\ell_{4}^{-2}=\ell^{-2}(1+\lambda)$, we notice the divergent contributions in the sum of the bulk and boundary actions precisely cancel. Subsequently taking the limit $\epsilon_{x,y}\to0^{+}$ gives
\beq
\begin{split} 
\hspace{-4mm} I_{E}^{\text{on-shell}}&=I_{\text{EH}}+I_{\text{Max}}+I_{\text{bdry}}\\
&=\frac{\ell^{2}\Delta t_{E}\Delta \phi}{8\pi G_{4}}\biggr[\left(\frac{(1+\lambda)}{(x_{1}-y_{+})^{2}}-\frac{1}{x_{1}^{2}}-\frac{\lambda}{y_{+}^{2}}\right)-\frac{a_{E}^{2}x_{1}^{2}y_{+}^{2}(1+\lambda)}{(x_{1}-y_{+})^{2}}-\frac{2q_{E}^{2}x_{1}y_{+}}{(1-a_{E}^{2}x_{1}^{2}y_{+}^{2})}\biggr]\;,
\end{split}
\label{eq:onshellactgeom}\eeq 
where we have multiplied by an overall factor of two because we are working with a $\mathbb{Z}_{2}$ construction. In the static and neutral limit we recover the on-shell action in \cite{Kudoh:2004ub,Panella:2024sor}.

\subsection{Quantum black hole thermodynamics}

With the Euclidean action in hand, we may now compute the thermodynamics using the grand canonical partition function $\mathcal{Z}(\beta,\Phi,\Omega)=-I_{E}^{\text{on-shell}}$,
\beq 
\begin{split}
&J= -\frac{1}{\beta}\left(\frac{\partial I_{E}^{\text{on-shell}}}{\partial\Omega}\right)_{\hspace{-1mm}\beta,\Phi}\;,\quad Q=-\frac{1}{\beta}\left(\frac{\partial I_{E}^{\text{on-shell}}}{\partial\Phi}\right)_{\hspace{-1mm}\beta,\Omega}\;,\\
&E=\left(\frac{\partial I_{E}^{\text{on-shell}}}{\partial\beta}\right)_{\hspace{-1mm}\Omega,\Phi}+\Omega J+\Phi Q\;,\quad S=\beta \left(\frac{\partial I_{E}^{\text{on-shell}}}{\partial\beta}\right)_{\hspace{-1mm}\Omega,\Phi}-I_{E}^{\text{on-shell}}\;,
\end{split}
\label{eq:thermoquantsgcens}\eeq
where $Q\equiv iQ_{E}$, $\Phi\equiv-i\Phi_{E}$, $J\equiv iJ_{E}$ and $\Omega\equiv -i\Omega_{E}$. In particular, the potentials are
\beq 
\begin{split}
&\beta=\frac{\ell\Sigma(x_{1},y_{+})}{(1+\bar{a}_{E}^{2}\ell^{2})\eta}\frac{4\pi}{H'(y_{+})}\;,\\
&\Phi=-\frac{2iq_{E}y_{+}\eta}{\ell_{\star}\Sigma(x_1,y_+)}\left(1+\bar{a}_{E}^{2}\ell^{2}\right)\equiv -i\Phi_{E}\;,\\
&\Omega=-i\frac{a_{E}}{\ell\Sigma(x_{1},y_{+})}\left(\frac{x_{1}^{2}\ell^{2}}{\ell_{3}^{2}}+y_{+}^{2}\right)\equiv -i\Omega_{E}\;.
\end{split}
\label{eq:potsyxco}\eeq
for real $Q_{E},\Phi_{E}, J_{E}$, and $\Omega_{E}$.

In principle, one can evaluate the derivatives (\ref{eq:thermoquantsgcens}) to compute the charges and entropy. For the current set-up, however, this is difficult in practice because $y_{+}$ and $x_{1}$ are related in a complicated way. To this end, it is convenient to introduce \cite{Emparan:2020znc,Climent:2024nuj,Panella:2024sor}
\beq z\equiv -\frac{y_+}{x_{1}}\;,\quad \tilde{q}\equiv qx_1^2\;,\quad \alpha_E\equiv \frac{a_{E}x_1}{\sqrt{k}}\;,\quad \nu\equiv \frac{\ell}{\ell_{3}}\;.\label{eq:zparam}\eeq
The parameters $x_1$, $y_+$ and $\mu$ can then be expressed as\footnote{The first expression follows from substituting the definition of $\mu$ (\ref{eq:muderiv}) and parameters (\ref{eq:zparam}) into $H(y_{+})=0$ and solving directly for $x_{1}^{2}$, from which the second relation directly follows via $y_{+}^{2}=x_{1}^{2}z^{2}$ and the third expression from employing (\ref{eq:muderiv}).}
\begin{align}\label{para ChRo}\notag
    x_1^2&=\frac{1}{k}\frac{\nu^2-z^3+z^3 \tilde q^2(1+z)}{z^2(1+z-\alpha_E^2 \nu^2 z+\alpha_E^2 z^2)}\;,\\
    y_+^2&=\frac{1}{k}\frac{\nu^2-z^3+z^3 \tilde q^2(1+z)}{1+z-\alpha_E^2 \nu^2z+\alpha_E^2 z^2}\;,\\ \notag
    \mu x_1&=\frac{k[(\nu^2+z^2)[1+\alpha_E^2(z^2- \nu^2)]+\tilde q^2 z^2(z^2-1)-\alpha_E^2\tilde q^2 (1+\nu^2)z^4]}{\nu^2-z^3+z^3 \tilde q^2(1+z)}\;.
\end{align}
Further rescaling $z \to \nu z$ and $\alpha_E \to \alpha_E/ \nu$ (to undo the coordinate change and parameter rescalings as in (\ref{eq:coordscalings}) and (\ref{eq:paramscals}), respectively), and subsequently replacing $k\to-\kappa$ and $\alpha_{E}=-i\alpha$, we recover the parameter expression Eq. (2.47) of \cite{Bhattacharya:2025tdn}.

Using this sequence of parameter rescalings, and $\Delta t_{E}=4\pi/H'(y_{+})$ and $\Delta\phi=2\pi \eta$, the on-shell action (\ref{eq:onshellactgeom}) becomes
\beq
\begin{split}
I^{\text{on-shell}}_{E}&= \frac{\pi\ell_{3}z\sqrt{1+\nu^{2}}}{G_{3}}\frac{\mathcal{N}}{\mathcal{D}}\;,      
\end{split}
\eeq
with numerator 
\beq 
\begin{split}
\hspace{-4mm} \mathcal{N}&=-1 + \alpha^4 \Big\{ 
    z (1 -z^2) \big( 2 z - \nu + z^3 (z - 2 \nu) \nu \big)- z^4 \nu \big(\nu - z (3 + 4 z \nu + z^3 \nu - \nu^2 - 3 z^2 (1 + \nu^2)) \big) \tilde{q}^2 \\
&+ z^8 \nu^2 (1 + \nu^2) \tilde{q}^4 
\Big\} -z \nu \Big[ 2 + z^2 \big( 2 + z \nu + 2\tilde{q}^{2} (1+ z \nu)^2 \big) \Big] \\
& \quad + \alpha^2 \Big\{ -1 + z \big( -3 \nu + z \big( 3 + z \nu \big( z (3 z + (-3 + z^2) \nu) \\
&- \big( 1 + z (6 \nu + z (-5 - 2 z \nu + (3 + z^2) \nu^2)) \tilde{q}^2 \big) \big) \big) \big) \Big\}\;,
\end{split}
\eeq
and denominator
\beq
\begin{split}
\hspace{-4mm} \mathcal{D}&= \Big[ 2 - z \Big( 4 z \alpha^2 - (3 + z^2 + (3 + z^4) \alpha^2) \nu+ z^2 \nu \big( 1 + 2 z \nu + z^2 (\nu^2 + \alpha^2 (1 + \nu^2)) \big) \tilde{q}^2 \Big) \Big] \\
& \times\Big[ 1 - \alpha^2 + z^2 \Big( 3 + \tilde{q}^2 - z \big( 3 z \alpha^2 + z \big( \alpha^2 - (1 - \alpha^2) \nu^2 \big) \tilde{q}^2 - 2 \nu (1 + 2 \alpha^2 + \tilde{q}^2) \big) \Big) \Big]\;.
\end{split}
\eeq
 Similarly, the potentials $\{\beta,\Omega,\Phi\}$ (\ref{eq:potsyxco})
 take the form
\beq
\begin{split}
 &\beta= \frac{2 \ell_{3} \pi z (1 + z \nu) \left( 1 + \alpha^2 \left( 1 - z^2 (1 - z \nu \tilde{q}^2) \right) \right)}{\alpha^2 - z^2 \left(1 + z \nu - z \alpha^2 \left( z - 2 \nu + \nu (1 + z \nu) \tilde{q}^2 \right) \right)}\\
&\times \frac{[-1 + \alpha^2 + z^2 \left( -3 - \tilde{q}^2 + z \left( 3 z \alpha^2 + z \left( \alpha^2 + (-1 + \alpha^2) \nu^2 \right) \tilde{q}^2 - 2 \nu (1 + 2 \alpha^2 + \tilde{q}^2) \right) \right)]}{[2 - z \left( 4 z \alpha^2 - (3 + z^2 + (3 + z^4) \alpha^2) \nu + z^2 \nu \left( 1 + 2 z \nu + z^2 (\nu^2 + \alpha^2 (1 + \nu^2)) \right) \tilde{q}^2 \right)]}\;,\\
&\Phi=\frac{4 z \nu\tilde{q} \left( 1 + z (\nu + \alpha^2 (\nu-z)) \right)}{\ell_{\star} (1 + z \nu) \left( 1 + \alpha^2 \left( 1 - z^2 (1 - z \nu \tilde{q}^2) \right) \right)}\\
&\times \frac{[\alpha^2 - z^2 \left(1 + z \nu - z \alpha^2 \left( z - 2 \nu + \nu (1 + z \nu) \tilde{q}^2 \right) \right)]}{[\alpha^2-1 -z^2 \left(3 + \tilde{q}^2 - z \left( 3 z \alpha^2 + z \left( \alpha^2 + (-1 + \alpha^2) \nu^2 \right) \tilde{q}^2 - 2 \nu (1 + 2 \alpha^2 + \tilde{q}^2) \right) \right)]}\;,\\
&\Omega=\frac{(1 + z^2) \alpha \sqrt{ \left( 1 + z (\nu + \alpha^2 (\nu-z)) \right) \left( 1 - z^3 \nu \left(1 - (1 + z \nu) \tilde{q}^2 \right) \right) }}{\ell_3 z (1 + z \nu) \left( 1 + \alpha^2 \left( 1 - z^2 (1 - z \nu \tilde{q}^2) \right) \right)}\;.
\end{split}
\eeq

While these quantities appear cumbersome, the computation of the thermodynamic variables, $E$, $S$, $Q$, and $J$ is now much easier to accomplish.
 Evaluating the constrained derivatives (\ref{eq:thermoquantsgcens}) (see Appendix \ref{app:cmetgeom} for further details), we find
\beq
\begin{split}
&E=\frac{\sqrt{1 + \nu^2} \, \left( 1 - z^3 \nu \left(1 - (1 + z \nu) \tilde{q}^2 \right) \right)}{2 G_3}\\
&\times \frac{[\alpha^2 + z^2 \Big( 1 + z \nu + 4 z \alpha^4 (\nu-z) + \alpha^2 \big( 4 + z ( 4 \nu-z + \nu (1 + z \nu) \tilde{q}^2) \big) \Big)]}{\Big[ 1 - \alpha^2 + z^2 \big( 3 + \tilde{q}^2 + 2 z \nu (1 + 2 \alpha^2 + \tilde{q}^2) + z^2 (\nu^2 \tilde{q}^2 - \alpha^2 (3 + (1 + \nu^2) \tilde{q}^2)) \big) \Big]^2}\;,\\
\end{split}
\eeq
\beq
\begin{split}
&J=\frac{\ell_3 z \alpha \sqrt{1 + \nu^2}}{G_{3}}\big( 1 + z^2 + \alpha^2 - z^4 \alpha^2 + z^3 (1 + \alpha^2) \nu (1 + z \nu) \tilde{q}^2 \big)\\
&\times \frac{\sqrt{\left( 1 + z (\nu + \alpha^2 (\nu-z)) \right) \left( 1 - z^3 \nu \left(1 - (1 + z \nu) \tilde{q}^2 \right) \right)}}{\Big[ 1 - \alpha^2 + z^2 \big( 3 + \tilde{q}^2 + 2 z \nu (1 + 2 \alpha^2 + \tilde{q}^2) + z^2 (\nu^2 \tilde{q}^2 - \alpha^2 (3 + (1 + \nu^2) \tilde{q}^2)) \big) \Big]^2}\;,
\end{split}
\eeq
\beq
\begin{split}
&Q=\frac{z^2 \sqrt{1 + \nu^2} \ell_\star \left( 1 + z (\nu + \alpha^2 (\nu - z)) \right) \tilde{q}}{2 G_3 \left[ 1 - \alpha^2 + z^2 \left( 3 + \tilde{q}^2 + 2 z \nu (1 + 2 \alpha^2 + \tilde{q}^2) + z^2 (\nu^2 \tilde{q}^2 - \alpha^2 (3 + (1 + \nu^2) \tilde{q}^2)) \right) \right]}\;.
\end{split}
\eeq
It is easy to verify the variables and conjugate potentials we derived match the thermodynamic quantities reported in \cite{Bhattacharya:2025tdn}. 

Further, written in terms of $(z,\alpha,\tilde{q})$, it can be verified that the on-shell action (\ref{eq:onshellactgeom}) has the expected form
\beq I_{E}^{\text{on-shell}}=\beta(E-\Phi Q-\Omega J)-S\;.\eeq
It is also straightforward to verify the thermodynamic quantities obey the first law
\beq \delta E=T\delta S+\Phi\delta Q+\Omega\delta J\;,\eeq
where we compute $(\partial_{z}E)_{\tilde{q},\alpha}=T(\partial_{z}S)_{\tilde{q},\alpha}+\Phi(\partial_{z}Q)_{\tilde{q},\alpha}+\Omega(\partial_{z}J)_{\tilde{q},\alpha}$ and similarly for derivatives with respect to $\tilde{q}$ and $\alpha$.

We can return to parameters $\{x_{1},y_{+},\ell,q_{E},a_{E}\}$ and show the thermodynamic entropy and charges $M,Q$, and $J$ are more compactly written as
\beq 
\begin{split}
&E=\frac{\ell k}{4G_{4}}\eta^{2}\left[1-\frac{a_{E}^{2}\ell^{2}x_{1}^{4}}{\ell_{3}^{2}}\left(1-\frac{4}{kx_{1}^{2}}\right)\right]=M\;,\\
&J=\frac{ia_{E}\ell^{2}}{2G_{4}}\eta^{2}\left(1+kx_{1}^{2}-\frac{a_{E}^{2}x_{1}^{4}\ell^{2}}{\ell_{3}^{2}}\right)\equiv iJ_{E}\;,\\
&Q=\frac{i q_{E} x_{1}\ell \ell_{\star}\eta}{2G_{4}}\equiv iQ_{E}\;,\\
&S=\frac{\pi \ell^{2}\eta x_{1}(1-a_{E}^{2}x_{1}^{2}y_{+}^{2})}{G_{4}y_{+}(y_{+}-x_{1})}=S^{(4)}_{\text{BH}}=S_{\text{gen}}^{(3)}\;,\\
\end{split}
\eeq
matching the mass (\ref{eq:massqbtz}), angular momentum (\ref{eq:angmomeqbtz}), electric charge (\ref{eq:elecchargeqBTZ}), and Bekenstein-Hawking area-entropy (\ref{eq:bulkent}).\footnote{The matching occurs by undoing the rotations and rescalings, $q\to iq_{E}$, $a\to ia_{E}\ell$, $\kappa=-k$, $y_{+}=-\ell/r_{+}$.} That the thermodynamic variables $\{E,J,Q\}$ and potentials $\{\Phi,\Omega\}$ are purely imaginary is standard for charged and rotating black holes.

In summary, we have established that the mass identified from the metric is indeed the (internal) thermal energy. Notably, from the brane perspective, when backreaction is accounted for, the classical gravitational entropy is supplanted by the generalized entropy in the first law of thermodynamics. This is consistent with the thermodynamics of exact two-dimensional quantum black holes \cite{Pedraza:2021cvx,Svesko:2022txo,Alexandre:2025hkr}.


\section{Charged and rotating black holes in dS$_{3}$ and Mink$_{3}$} \label{sec:bhsindSMink3}

In (2+1)-dimensional vacuum general relativity, the only known example of a black hole is the BTZ solution \cite{Banados:1992gq,Banados:1992wn}. Massive point particles in (2+1)-dimensional Minkowski (Mink$_{3}$) or de Sitter (dS$_{3}$) are conical defects with a conical singularity at the origin \cite{Deser:1983tn,Deser:1983nh}; in the case of the Schwarzschild- or Kerr-dS$_{3}$ there is only a single \emph{cosmological} horizon. The reason for the presence of black holes in AdS$_{3}$ is that the negatively curved geometry leads to a natural mechanism for gravitational gravitational collapse \cite{Ross:1992ba}. Black holes in three-dimensions can arise when another scale is introduced. For example, dS$_{3}$ black holes appear in pure modified theories of gravity, e.g., `new massive gravity' \cite{deBuyl:2013ega} or topological massive gravity \cite{Nutku:1993eb,Anninos:2009jt}. Quantum effects also facilitate the generation of (quantum) black holes in dS$_{3}$ and Mink$_{3}$ \cite{Emparan:2022ijy,Panella:2023lsi,Climent:2024wol}, where semi-classical backreaction modifies the geometry so as to induce black hole horizons (and curvature singularities) where there were none before.  

Below we use holographic braneworlds to construct charged and rotating quantum black holes in dS$_{3}$ and Mink$_{3}$, characterizing both their geometry and thermal description. As we describe below, this is accomplished by introducing Randall-Sundrum ETW branes \cite{Randall:1999ee,Randall:1999vf}, where the brane has de Sitter or flat asymptotics depending on the brane tension.

\subsection{Black holes in dS$_{3}$}

\subsubsection*{Geometry}

\noindent \textbf{Bulk and brane set-up.} As before, we again start with Einstein-Maxwell-AdS$_{4}$ gravity, and the charged and rotating C-metric (\ref{eq:rotatingCmetqbtz}), now with metric functions 
\beq
\begin{split}
    &H(r)= 1-\frac{r^2}{R_3^2} -\frac{\mu \ell}{r}+ \frac{a^2}{r^2}+\frac{q^{2}\ell^{2}}{r^{2}}  \ , \quad G(x)= 1- x^2-\mu x^3- \frac{a^2}{R_3^2}x^4-q^{2}x^{4} \ , \\
   &\Sigma(x,r)= 1 + \frac{a^2 x^2}{r^2} \ ,
\end{split}
\label{eq:metfuncsrotqdS}\eeq
where we fixed $\kappa=+1$, send $a\to-a$ (out of convention) and set $\ell_{3}^{2}=-R_{3}^{2}$ (or $\ell_{3}\to iR_{3}$). The AdS$_{4}$ length scale is now
\beq \frac{1}{\ell_{4}^{2}}=\frac{1}{R_{3}^{2}}-\frac{1}{\ell^{2}}\;.\label{eq:L4bulkRS}\eeq
Thence, maintaining $\ell_{4}^{2}>0$ requires $R_{3}^{2}>\ell^{2}$, placing an upper bound on $\ell$ for fixed $R_{3}$, and $\ell<\ell_{4}$. Consequently, unlike the qBTZ  construction, we are outside of the regime of slow acceleration  where $\ell>\ell_{4}$; the acceleration horizon will imprint on the brane.

As in the qBTZ black hole, real roots of $G(x)$ correspond to orbits of $\partial_{\phi}$ and characterize the shape of the horizon.  We again restrict to the range $0\leq x\leq x_{1}$, for smallest root of $G(x)=0$, $x_{1}$, where the conical singularity at $x=x_{1}$ is removed via the identification (\ref{eq:azimuthid}), $\phi\sim \phi+\Delta\phi$, with 
\beq \Delta\phi=\frac{4\pi}{|G'(x_{1})|}=4\pi\biggr|-2 x_{1}-3\mu x_{1}^{2}-4\left(q^{2}+\frac{a^{2}}{R^{2}_{3}}\right)x_{1}^{3}\biggr|^{-1}\;.\eeq
Further, solving $G(x_{1})=0$, we write the mass parameter as
\beq \mu=\frac{1}{x_{1}^{3}}\left[1-x_{1}^{2}-\left(q^{2}+\frac{a^{2}}{R_{3}^{2}}\right)x_{1}^{4}\right]\;,\label{eq:muderivdS}\eeq
which is taken to be positive. 

The hypersurface at $x=0$ remains totally umbilic, where again  $K_{ij}=-\ell^{-1}h_{ij}$. Placing an ETW brane there obeying Israel junction conditions (\ref{eq:israeljuncconds}) sets the tension to (\ref{eq:branetengen}). The induced geometry at $x=0$ is 
\beq ds^{2}|_{x=0}=-H(r)dt^{2}+H^{-1}(r)dr^{2}+r^{2}\left(d\phi+\frac{a}{r}dt\right)^{2}\;. \label{eq:naivemetrotds}\eeq
An important effect of the brane is that its intersection with the non-compact bulk acceleration horizon  (as the horizon extends to the asymptotic boundary), induces a compact horizon understood to be the dS$_{3}$ cosmological horizon. We will return to this point momentarily. 

Analogous to coordinates (\ref{eq:transKillvec}), the canonically normalized coordinates are \cite{Panella:2023lsi}
\beq t=\eta (\bar{t}+\bar{a}R_{3}\bar{\phi})\;,\quad r=\sqrt{\frac{\bar{r}^{2}-r_{s}^{2}}{(1+\bar{a}^{2})\eta^{2}}}\;,\quad \phi=\eta\left(\bar{\phi}-\frac{\bar{a}\bar{t}}{R_{3}}\right)\;,\label{eq:transKillvecdS}\eeq
for
\beq \eta\equiv \frac{\Delta\phi}{2\pi}\;,\quad \bar{a}\equiv \frac{ax_{1}^{2}}{R_{3}}\;,\quad r_{s}\equiv -\frac{R_{3}\bar{a}\eta}{x_{1}}\sqrt{2- x_{1}^{2}}\;.\label{eq:transparadS}\eeq
The induced metric at $x=0$ is then 
\beq 
\begin{split}
ds^{2}|_{x=0}&=-\left[-\frac{\bar{r}^{2}}{R_{3}^{2}}-\eta^{2}\left(-\frac{4\bar{a}^{2}}{x_{1}^{2}}-(1-\bar{a}^{2})\right)-\frac{\mu\ell\eta^{2}}{r}+\frac{q^{2}\ell^{2}\eta^{2}}{r^{2}}\right]d\bar{t}^{2}\\
&-\frac{\bar{r}^{2}}{(1+\bar{a}^{2})\eta^{4}r^{2}}\left[1-\frac{r^{2}}{R_{3}^{2}}-\frac{\mu\ell}{r}+\frac{q^{2}\ell^{2}x_{1}^{2}+\bar{a}^{2}R_{3}^{2}}{x_{1}^{4}r^{2}}\right]^{-1}d\bar{r}^{2}\\
&+\left[\bar{r}^{2}-\bar{a}^{2}\eta^{2}R_{3}^{2}\left(\frac{q^{2}\ell^{2}}{r^{2}}-\frac{\mu\ell}{r}\right)\right]d\bar{\phi}^{2}\\
&+2\bar{a}x_{1}R_{3}\eta^{2}(\mu+q^{2}x_{1})\left[1+\frac{\ell}{x_{1}}-\frac{q^{2}\ell(x_{1}r+\ell)}{r^{2}x_{1}(q^{2}x_{1}+\mu)}\right]d\bar{t}d\bar{\phi}\;.
\end{split}
\label{eq:cannormmetdS}\eeq
We recover the induced asymptotically AdS$_{3}$ metric (\ref{eq:cannormmet}) upon sending $\ell_{3}\to iR_{3}$, and $\bar{a}_{\text{AdS}_{3}} \to i\bar{a}_{\text{dS}_{3}}$.  Further, the gauge potential  in canonically normalized coordinates on the brane has the form
\beq \bar{A}_{b}d\bar{x}^{b}|_{x=0}=-\frac{2q\ell\eta}{\ell_{\star}r(\bar{r})}d\bar{t}-\frac{2\bar{a}q\ell R_{3}\eta}{\ell_{\star}r(\bar{r})}d\bar{\phi}\;.\label{eq:gaugepotqdS}\eeq

\vspace{2mm}

 \noindent \textbf{Horizon bestiary.} Unlike the Karch-Randall construction, the Randall-Sundrum ETW brane at $x=0$ intersects both the bulk black hole and acceleration horizons. Consequently, the braneworld geometry (in naive or normalized coordinates) retains the same inner and outer black hole horizons and a compact cosmological horizon. Indeed, horizons in the bulk correspond to positive real roots $r_{i}$ of $H(r)$.
To classify the types of horizons, it is useful to define  $Q(r)\equiv r^{2}H(r)$, a quartic polynomial in $r$ with four, two, or zero distinct real roots. We will focus on the case of four real roots, three of which will be positive, corresponding to the cosmological horizon $r_{c}$ and outer/inner black hole horizons $r_{\pm}$, obeying $r_{-}\leq r_{+}\leq r_{c}$; the fourth root, $r_{n}=-(r_{c}+r_{+}+r_{-})<0$. We can express 
\beq
\begin{split}
&R_{3}^{2}=r_{c}^{2}+r_{+}^{2}+r_{c}r_{+}+r_{-}(r_{c}+r_{+}+r_{-})\;,\\
&\mu\ell=\frac{(r_{c}+r_{+})(r_{c}+r_{-})(r_{+}+r_{-})}{r_{c}^{2}+r_{+}^{2}+r_{c}r_{+}+r_{-}(r_{c}+r_{+}+r_{-})}\;,\\
&a^{2}+q^{2}\ell^{2}=\frac{r_{c}r_{+}r_{-}(r_{c}+r_{+}+r_{-})}{r_{c}^{2}+r_{+}^{2}+r_{c}r_{+}+r_{-}(r_{c}+r_{+}+r_{-})}\;.
\end{split}
\label{eq:paramsintermsofri}\eeq
where we used $H(r_{c})=0$, and $H(r_{\pm})=0$.

In canonically normalized coordinates the Killing horizon generator is (\ref{eq:cannormgenerator})
with black hole horizon angular velocity 
\beq \bar{\Omega}_{\pm}=-\frac{a}{r_{\pm}^{2}+a^{2}x_{1}^{2}}\left(1-\frac{r_{\pm}^{2}x_{1}^{2}}{R_{3}^{2}}\right)\;.\label{eq:horangveldS}\eeq
Moreover, the surface gravity associated with each horizon is given by 
\beq \bar{\kappa}_{i}=\frac{\eta(1+\bar{a}^{2})}{\left(r_{i}^{2}+a^{2}x_{1}^{2}\right)}\frac{r_{i}^{2}}{2}|H'(r_{i})|=\frac{\eta(1+\bar{a}^{2})}{\left(r_{i}^{2}+a^{2}x_{1}^{2}\right)}\frac{1}{2R_{3}^{2}r_{i}}|R_{3}^{2}\mu\ell r_{i}-2r_{i}^{4}-2(a^{2}+q^{2}\ell^{2})R_{3}^{2}|
\;.\label{eq:surfacegravsdS}\eeq
Using the parameters (\ref{eq:paramsintermsofri}), we explicitly have 
\beq 
\begin{split}
&\bar{\kappa}_{c}=-\frac{\eta(1+\bar{a}^{2})}{2R_{3}^{2}\left(r_{c}^{2}+a^{2}x_{1}^{2}\right)}(r_{c}-r_{+})(r_{c}-r_{-})(r_{+}+r_{-}+2r_{c})\;,\\
&\bar{\kappa}_{+}=\frac{\eta(1+\bar{a}^{2})}{2R_{3}^{2}\left(r_{+}^{2}+a^{2}x_{1}^{2}\right)}(r_{c}-r_{+})(r_{+}-r_{-})(r_{c}+r_{-}+2r_{+})\;,\\
&\bar{\kappa}_{-}=-\frac{\eta(1+\bar{a}^{2})}{2R_{3}^{2}\left(r_{-}^{2}+a^{2}x_{1}^{2}\right)}(r_{c}-r_{-})(r_{+}-r_{-})(r_{c}+r_{+}+2r_{-})\;,
\end{split}
\label{eq:surfacegravsv2}\eeq
for cosmological horizon surface gravity $\bar{\kappa}_{c}$ and black hole horizon surface gravities $\bar{\kappa}_{\pm}$.

\vspace{2mm}

 \noindent \textbf{Quantum black holes.} As in the quantum BTZ construction, there is an induced theory of semi-classical gravity on the brane. The induced couplings of the brane theory include Newton's three-dimensional constant, the Maxwell couplings (\ref{eq:effLd}), and the effective dS$_{3}$ cosmological constant \cite{Emparan:2022ijy},
 \beq \Lambda_{3}=\frac{2}{L_{3}^{2}}\equiv \frac{4}{\ell_{4}}\left(\frac{1}{\ell}-\frac{1}{\ell_{4}}\right)\;,\eeq
 where $\ell_{4}\ll L_{3}$ and $\ell\sim\ell_{4}\ll R_{3}$. Further, the central charge $c_{3}$ of the $\text{CFT}_{3}$ is 
\beq c_{3}=\frac{\ell_{4}^{2}}{G_{4}}=\frac{\ell}{2G_{3}\sqrt{1-(\ell/R_{3})^{2}}}\;.\eeq
The semi-classical theory has field equations (\ref{eq:semiclasseombrane}) and (\ref{eq:sccurrentdens}), except where $L_{3}\to iL_{3}$.

From the brane perspective, the induced metric (\ref{eq:cannormmetdS}) can be expressed as
\beq
\begin{split}
    ds^2=&-\left(1-8\mathcal{G}_3M-\frac{\bar r^2}{R_3^2}-\frac{\mu \ell \eta^2}{r}+\frac{q^2 \ell^2\eta^2}{r^2}\right)d\bar t^2\\
    &+\left(1-8\mathcal{G}_3 M -\frac{\bar r^2}{R_3^2}+\frac{(4\mathcal{G}_3 J)^2}{\bar r^2}+\frac{(1+\bar{a}^{2})^{2}\eta^{4}}{\bar{r}^{2}}(q^{2}\ell^{2}-\mu\ell r) \right)^{-1}d\bar r^2\\ 
    &+\left(\bar{r}^{2}+\bar{a}^{2}R_{3}^{2}\ell\eta^{2}\left(\frac{\mu}{r}-\frac{q^{2}\ell}{r^{2}}\right)\right)d\bar{\phi}^{2}-8\mathcal{G}_{3}J\left(1+\frac{\ell}{x_{1}r}-\frac{q^{2}\ell(x_{1}r+\ell)}{r^{2}x_{1}(q^{2}x_{1}+\mu)}\right)d\bar{t}d\bar{\phi}\;,
\end{split}
\label{eq:qKNdSmet}\eeq
where we have identified the mass $M$ and angular momentum $J$ as
\beq 
\begin{split}
&1-8\mathcal{G}_{3}M\equiv \eta^{2}\left(1-\bar{a}^{2}+\frac{4\bar{a}^{2}}{x_{1}^{2}}\right)\;,\\
&4\mathcal{G}_{3}J\equiv-\bar{a}R_{3}x_{1}\eta^{2}(\mu+x_{1}q^{2})\;.
\end{split}
\eeq
Additionally, the electric charge is given by (\ref{eq:elecchargeqBTZ}), while the conjugate electric potential is 
\beq \Phi\equiv\frac{2\ell q\eta}{\ell_{\star}r_{+}\Sigma(x_{1},r_{+})}(1+\bar{a}^{2})\;.\label{eq:potentialqKN}\eeq
In the vanishing charge limit $q\to0$, one recovers the neutral rotating quantum Kerr-dS$_{3}$ black hole \cite{Panella:2023lsi}, while in the limit of vanishing angular momentum we recover the charged quantum dS$_{3}$ black hole \cite{Climent:2024wol}.

We can compute the components of the quantum matter stress-tensor using the perturbative expansion of the field equations (\ref{eq:Tij0gen}) and (\ref{eq:Tij2gen}). Here we simply report the leading contribution in canonically normalized coordinates: 
\begin{align}\notag
      \langle T\indices{^{\bar{t}}_{\bar {t}}}\rangle_0&=\frac{\ell}{16\pi G_3(1+\bar a^2)r^3} \left[\mu\left(1-2\bar a^2+\frac{3 \bar a^2 R_3^2}{x_1^2 r^2}\right)-\frac{q^2\ell}{r}\left(\frac{1+7 \bar a^2}{2}+\frac{3\bar a^2 R_3^2 }{x_1^2r^2}\right)\right]\;,\\ \notag
        \langle T\indices{^{\bar{r}}_{\bar{r}}}\rangle_0&=\frac{2\mu \ell r+q^2 \ell^2}{32\pi G_3 r^4}\;,\\
          \langle T\indices{^{\bar{\phi}}_{\bar{\phi}}}\rangle_0&=-\frac{\ell}{16\pi G_3(1+\bar a^2)r^3}\left[\mu\left(2-\bar a^2+\frac{3\bar a^2 R_3^2}{x_1^2 r^2}\right)+\frac{q^2 \ell}{r}\left(\frac{7+\bar a^2}{2}-\frac{3\bar a^2 R_3^2 }{x_1^2 r^2}\right)\right]\;,\\ \notag
            \langle T\indices{^{\bar{t}}_{\bar{\phi}}}\rangle_0&=\frac{3 \ell \bar a R_3}{16\pi G_3(1+\bar a^2)r^3}\left(\mu-\frac{q^2\ell}{r}\right)\left(1+\frac{\bar a^2 R_3^2}{x_1^2 r^2}\right)\;,\\ \notag
            \langle T\indices{^{\bar{\phi}}_{\bar{t}}}\rangle_0&=\frac{3 \ell \bar a}{16\pi G_3(1+\bar a^2)R_3 r^3}\left(\mu-\frac{q^2\ell}{r}\right)\left(1-\frac{R_3^2}{x_1^2 r^2}\right)\;.
\end{align}
In the neutral limit, $q\to0$, the quantum stress-tensor components reduce to those supporting the quantum Kerr-dS$_{3}$ black hole \cite{Panella:2023lsi}.

\subsubsection*{Thermodynamics}

Due to the presence of the acceleration horizon in the bulk, quantum black holes in dS$_{3}$ generically have at least two horizons, a cosmological horizon and a black hole horizon. As noted above, these horizons generally have different surface gravities. Consequently, to a static patch observer who experiences both horizons, a de Sitter black hole will generically not be in thermal equilibrium. This makes evaluating the thermodynamics of dS black holes more subtle than their flat or AdS counterparts. 

Indeed, deriving the horizon thermodynamics for dS black holes from an action integral is more intricate. Applying the standard Gibbons-Hawking recipe, the on-shell Euclidean action of empty dS space yields (minus) the gravitational entropy of the cosmological horizon. When embedding a black hole inside an asymptotically dS background, however, the standard recipe for evaluating the on-shell action is not possible. For example, the Euclidean section of Schwarzschild-dS (SdS) spacetime generically has two conical singularities, for which only one is eliminated by identifying the inverse periodicity of the Euclidean time circle with the black hole or cosmological horizon temperature. Despite the geometry being singular, the on-shell action of the (classical) SdS solution is finite and equal to (minus) the sum of the entropies of the two horizons \cite{Chao:1997osu,Bousso:1998na,Gregory:2013hja,Morvan:2022ybp,Morvan:2022aon}.

Likewise, although the Euclidean AdS$_{4}$ C-metric geometry considered here is generically singular (due to the acceleration horizon), its on-shell action is finite. Indeed, its action is given by (\ref{eq:onshellactgeom}) for $-1<\lambda<0$. Consequently, the thermodynamic potentials for the rotating and charged quantum dS$_{3}$ black hole are
\beq
\begin{split}
 &\beta_{i}=
\frac{
2 \pi R_3 \, z \, (1 + z \nu)\left(1 + \alpha^2 \left(1 + z^2 \left(1 - z \nu \tilde{q}^2\right)\right)\right)
}{
\alpha^2 + z^2 \left(1 + z \left(\nu + \alpha^2 (z + 2\nu)\right) - z \alpha^2 \nu (1 + z \nu)\tilde{q}^2\right)
}\\
&\times \frac{
1 - \alpha^2 - z^2 \left(3 + \tilde{q}^2\right)
- 2 z^3 \nu \left(1 + 2 \alpha^2 + \tilde{q}^2\right)
+ z^4 \left(-\nu^2 \tilde{q}^2 + \alpha^2 \left(-3 + (-1 + \nu^2)\tilde{q}^2\right)\right)
}{
\biggr|2 + z \Bigl(
4 z \alpha^2 + 3 \nu - z^2 \nu + 3 \alpha^2 \nu + z^4 \alpha^2 \nu
+ z^2 \nu \left(1 + 2 z \nu + z^2 \left(\nu^2 + \alpha^2 (-1 + \nu^2)\right)\right)\tilde{q}^2
\Bigr)\biggr|
}\;,\\
&\Phi_{i}=
-\frac{
4 z \nu \left(1 + z\left(\nu + \alpha^2 (z + \nu)\right)\right)\tilde{q}
}{
\ell_{\star}\,(1 + z \nu)\left(1 + \alpha^2 \left(1 + z^2 \left(1 - z \nu \tilde{q}^2\right)\right)\right)
}\\
&\times \frac{
[\alpha^2
+ z^2 \left(1 + z \left(\nu + \alpha^2 (z + 2\nu)\right)
+ z \alpha^2 \nu (1 + z \nu)\tilde{q}^2 \right)
]}{[
1 - \alpha^2
- z^2 \left(3 + \tilde{q}^2\right)
- 2 z^3 \nu \left(1 + 2 \alpha^2 + \tilde{q}^2\right)
- z^4 \left(\nu^2 \tilde{q}^2 + \alpha^2 \left(3 + (1 - \nu^2)\tilde{q}^2\right)\right)
]}\;,\\
&\Omega_{i}=\frac{
\left(z^2 - 1\right)\,\alpha\,
\sqrt{
\left(1 + z\left(\nu + \alpha^2 (z + \nu)\right)\right)
\left(1 + z^3 \nu \left(1 - (1 + z \nu)\tilde{q}^2\right)\right)
}
}{
R_3\, z \,(1 + z \nu)\left(1 + \alpha^2 \left(1 + z^2 \left(1 - z \nu \tilde{q}^2\right)\right)\right)
}\;,
\end{split}
\eeq
where here\footnote{More carefully, here by $z$ we mean $z_{i}$, to distinguish each horizon $r_{i}$. For notational simplicity we use only $z$, unless otherwise explicitly specified.} 
\begin{equation}\label{paradS}
    z=\frac{R_3}{r_i x_1}\;,\quad \nu=\frac{\ell}{R_3}\;,\quad \alpha=\frac{ax_{1}}{R_{3}}\;, 
\end{equation}
such that 
\beq 
\begin{split}
&x_{1}^{2}=
\frac{
1 + z^3 \nu \left(1 - (1 + z \nu)\tilde{q}^2\right)
}{
z^2 \left(1+ z  \left(\nu + \alpha^2 (z + \nu)\right)\right)
}\;,\\
&r_{i}^{2}=
\frac{
R_3^2 \left[1 + z\left(\nu + \alpha^2 (z + \nu)\right)\right]
}{
1 + z^3 \nu \left(1 - (1 + z \nu)\tilde{q}^2\right)
}\;,\\
&\mu x_{1}=
-\,\frac{
1 + \alpha^2
- z^2 \left(1 - \tilde{q}^2\right)
- z^4 \left(\alpha^2 + \left(\nu^2 + \alpha^2 (\nu^2 - 1)\right)\tilde{q}^2\right)
}{
1 + z^3 \nu \left(1 - (1 + z \nu)\tilde{q}^2\right)
}\;,
\end{split}
\eeq
for horizon radii $\{r_{i}\}=\{r_{\pm},r_{c}\}$. 

In terms of these parameters, the thermodynamic variables are
\beq
\begin{split}
&E=\frac{
\sqrt{1 - \nu^2} \,\left(1 + z^3 \nu \left(1 - (1 + z \nu)\tilde{q}^2\right)\right)
}{
2 G_3
}\\
&\times
\frac{
\alpha^2 - z^2 \Bigl(1 + z \nu 
+ 4 z \alpha^4 (z + \nu) 
+ \alpha^2 \bigl(4 + z (z + 4 \nu + \nu (1 + z \nu) \tilde{q}^2)\bigr)\Bigr)
}{
\Bigl[1 - \alpha^2 - z^2 \Bigl(3 + \tilde{q}^2 
+ z \bigl(2 \nu + \alpha^2 (3 z + 4 \nu) + (2 \nu + z (\nu^2 - \alpha^2 (-1 + \nu^2))) \tilde{q}^2\bigr)\Bigr)\Bigr]^2
}\;,
\end{split}
\eeq
\beq
\begin{split}
&J=
\frac{
R_3 z \alpha \sqrt{1 - \nu^2}
}{G_3}\left[z^2 - 1 + (z^4 - 1) \alpha^2 + z^3 (1 + \alpha^2) \nu (1 + z \nu) \tilde{q}^2 \right]\\
&\times 
\frac{
\sqrt{
\left(1 + z \left(\nu + \alpha^2 (z + \nu)\right)\right)
\left(1 + z^3 \nu \left(1 - (1 + z \nu)\tilde{q}^2\right)\right)
}
}{
\left[
1 - \alpha^2 - z^2 \Bigl(
3 + \tilde{q}^2 + z \bigl(2 \nu + \alpha^2 (3 z + 4 \nu) + (2 \nu + z (\nu^2 - \alpha^2(-1 + \nu^2))) \tilde{q}^2\bigr)
\Bigr)
\right]^{2}
}\;,
\end{split}
\eeq
\beq
\begin{split}
&Q=
- \frac{
z^2 \sqrt{1 - \nu^2} \left(1 + z (\nu + \alpha^2 (z + \nu))\right) \tilde{q} \ell_{\star}
}{
2 G_3 \left[
1 - \alpha^2 - z^2 (3 + \tilde{q}^2) 
- 2 z^3 \nu (1 + 2 \alpha^2 + \tilde{q}^2) 
- z^4 \bigl(\nu^2 \tilde{q}^2 + \alpha^2 (3 + (1 - \nu^2) \tilde{q}^2)\bigr)
\right]
}
\;.
\end{split}
\eeq
Upon setting $\tilde{q}=0$, we recover the thermodynamic quantities for the quantum Kerr-dS$_{3}$ black hole \cite{Panella:2023lsi}, while for $\alpha=0$ we recover charged quantum SdS$_{3}$ thermodynamics. 

Using (\ref{paradS}) we find the internal energy equals the mass $M$ of the black hole, while, from the brane perspective $S=S_{\text{gen}}^{(3)}$. Additionally, by explicit computation, we find the above thermodynamic quantities obey the first law
\beq \delta E=T_{i}\delta S+\Phi_{i}\delta Q+\Omega_{i}\delta J\;,\eeq
where, $T_{c}=-T_{i}(z_{c})$, $T_{+}=T_{i}(z_{+})$, and $T_{-}=-T_{i}(z_{-})$, while $\Omega_{c}=\Omega_{i}(z_{c})$, $\Omega_{\pm}=\Omega_{i}(z_{\pm})$, and $\Phi_{c}=\Phi_{i}(z_{c})$ and $\Phi_{\pm}=\Phi_{i}(z_{\pm})$. Thus, we have a semi-classical first law for each horizon,
\beq
\begin{split}
&\delta M=T_{+}\delta S^{(3)}_{\text{gen}}+\Phi_{+}\delta Q+\Omega_{+}\delta J\;,\\
&\delta M=-T_{-}\delta S^{(3)}_{\text{gen}}+\Phi_{-}\delta Q+\Omega_{-}\delta J\;,\\
&\delta M=-T_{c}\delta S^{(3)}_{\text{gen}}+\Phi_{c}\delta Q+\Omega_{c}\delta J\;.
\end{split}
\eeq
We note the minus sign in the first law of the cosmological horizon. Hence, as for classical dS horizons, adding positive energy into the static patch reduces the total entropy of the system, implying the entropy of pure dS
is maximal.

\subsection{Black holes in Mink$_{3}$}

\subsubsection*{Geometry}

\noindent \textbf{Bulk and brane set-up.} We start again with the C-metric \eqref{eq:rotatingCmetqbtz}, now with metric functions
\begin{equation}
    \begin{split}
    &H(r)=1 -\frac{\mu \ell}{r}+ \frac{a^2}{r^2}+\frac{q^{2}\ell^{2}}{r^{2}}  \ , \quad G(x)= 1-x^2-\mu x^3+q^{2}x^{4} \ , \\
   &\Sigma(x,r)= 1 + \frac{a^2 x^2}{r^2} \ . 
\end{split}
\end{equation}
where we have set $\kappa=+1$. It is easy to verify the bulk geometry is an Einstein metric with a negative cosmological constant with AdS$_{4}$ length scale $\ell_{4}=\ell$.
This means we are not working in the slow acceleration regime, however, in our construction the acceleration horizon will not intersect with the ETW brane.  

Real zeros of $G(x)$ of course distort the horizon and generate conical singularities. As usual, we restrict to the range $0\leq x\leq x_{1}$, removing the conical singularity at $x=x_{1}$ via the coordinate identification $\phi\sim \phi+\Delta\phi$ with 
\beq \Delta\phi=\frac{4\pi}{|G'(x_{1})|}=4\pi\biggr|-2 x_{1}-3\mu x_{1}^{2}-4q^{2}x_{1}^{3}\biggr|^{-1}\;,\quad \mu=\frac{1}{x_{1}^{3}}\left[1-x_{1}^{2}-q^{2}x_{1}^{4}\right]\;,\label{eq:muderivdS}\eeq
with $\mu\geq0$. The hypersurface at $x=0$ remains umbilic, such that the Israel junction conditions for an ETW brane sets the tension to (\ref{eq:branetengen}). 

\vspace{2mm}

\noindent \textbf{Quantum black holes.} The induced geometry at $x=0$ has the same form as (\ref{eq:naivemetrotds}). In terms of canonically normalized \cite{Panella:2024sor}
\beq t=\eta (\bar{t}+ax_{1}^{2}\bar{\phi})\;,\quad r=\sqrt{\frac{\bar{r}^{2}-r_{s}^{2}}{\eta^{2}}}\;,\quad \phi=\eta\bar{\phi}\;,\quad r_{s}\equiv -ax_{1}\eta\sqrt{2- x_{1}^{2}}\;,\label{eq:transKillvecflat}\eeq
the induced metric at $x=0$ can be cast as

\beq
\begin{split}
    ds^2=&-\left(1-8\mathcal{G}_3M-\frac{\mu \ell \eta^2}{r}+\frac{q^2 \ell^2\eta^2}{r^2}\right)d\bar t^2\\
    &+\left(1-8\mathcal{G}_3 M+\frac{(4\mathcal{G}_3 J)^2}{\bar r^2}+\frac{\eta^{4}}{\bar{r}^{2}}(q^{2}\ell^{2}-\mu\ell r) \right)^{-1}d\bar r^2\\ 
    &+\left(\bar{r}^{2}+a^{2}x_{1}^{4}\ell\eta^{2}\left(\frac{\mu}{r}-\frac{q^{2}\ell}{r^{2}}\right)\right)d\bar{\phi}^{2}-8\mathcal{G}_{3}J\left(1+\frac{\ell}{x_{1}r}-\frac{q^{2}\ell(x_{1}r+\ell)}{r^{2}x_{1}(q^{2}x_{1}+\mu)}\right)d\bar{t}d\bar{\phi}\;,
\end{split}
\label{eq:qKNflmet}\eeq
where we have identified the mass $M$ and angular momentum $J$ as
\beq 
\begin{split}
&1-8\mathcal{G}_{3}M\equiv \eta^{2}\;,\\
&4\mathcal{G}_{3}J=-ax_{1}^{3}\eta^{2}(\mu+x_{1}q^{2})\;.
\end{split}
\label{eq:massflat}\eeq
 The gauge potential on the brane is (\ref{eq:gaugepotqdS}), supporting the electric charge  (\ref{eq:elecchargeqBTZ}), with conjugate electric potential
\beq \Phi\equiv\frac{2\ell q\eta}{\ell_{\star}r_{+}\Sigma(x_{1},r_{+})}\;.\label{eq:potentialqKNfl}\eeq
Note that here $\mathcal{G}_{3}=G_{3}$.

 Unlike the above construction with dS$_{3}$ branes, the  acceleration horizon sits at $r\to \infty$ and does not intersect the brane (except at single point on the AdS$_{4}$ boundary), corresponding to the bulk AdS$_{4}$ horizon. See the discussion in Section 4.4 of \cite{Panella:2024sor} for more details. As such, the geometry localized on the brane will have only an inner and outer horizon $r_{\pm}$, real positive roots of $H(r_{\pm})=0$. In canonically normalized coordinates the horizon generator is (\ref{eq:cannormgenerator}) for black hole horizon angular velocity measured at spatial infinity
\beq \bar{\Omega}_{\pm}=-\frac{a}{r_{\pm}^{2}+a^{2}x_{1}^{2}}\;.\label{eq:horangvelflat}\eeq
Further, the surface gravity associated with each horizon is given by 
\beq \bar{\kappa}_{\pm}=\frac{\eta}{\left(r_{\pm}^{2}+a^{2}x_{1}^{2}\right)}\frac{r_{\pm}^{2}}{2}|H'(r_{\pm})|=\frac{\eta}{\left(r_{\pm}^{2}+a^{2}x_{1}^{2}\right)}\frac{1}{2r_{i}}|\mu\ell r_{\pm}-2(a^{2}+q^{2}\ell^{2})|
\;.\label{eq:surfacegravsflat}\eeq

From the brane perspective, since $\ell=\ell_{4}$,  the geometry (\ref{eq:qKNflmet}) describes a rotating and charged quantum black hole in Mink$_{3}$, supported by a CFT$_{3}$ with central charge $c_{3}=\frac{\ell}{2G_{3}}$. It is an exact solution to the semi-classical field equations (\ref{eq:semiclasseombrane}) and (\ref{eq:sccurrentdens}) (with $L_{3}\to\infty$).  Using the perturbative expansion of the field equations (\ref{eq:Tij0gen}), the leading contributions to the quantum matter stress tensor in an small-$\ell$ expansion are
\begin{align}\notag
      \langle T\indices{^{\bar{t}}_{\bar {t}}}\rangle_0&=\frac{\ell}{16\pi G_3 r^3} \left[\mu\left(1+\frac{3 a^{2}x_{1}^{2}}{ r^2}\right)-\frac{q^2\ell}{r}\left(\frac{1}{2}+\frac{3a^2 x_{1}^{2}}{r^2}\right)\right]\;,\\ \notag
        \langle T\indices{^{\bar{r}}_{\bar{r}}}\rangle_0&=\frac{2\mu \ell r+q^2 \ell^2}{32\pi G_3 r^4}\;,\\
          \langle T\indices{^{\bar{\phi}}_{\bar{\phi}}}\rangle_0&=-\frac{\ell}{16\pi G_3 r^3}\left[\mu\left(2+\frac{3 a^2 x_{1}^{2}}{ r^2}\right)+\frac{q^2 \ell}{r}\left(\frac{7}{2}-\frac{3 a^2 x_{1}^{2}}{r^2}\right)\right]\;,\\ \notag
            \langle T\indices{^{\bar{t}}_{\bar{\phi}}}\rangle_0&=\frac{3 \ell a x_{1}^{2}}{16\pi G_3 r^3}\left(\mu-\frac{q^2\ell}{r}\right)\left(1+\frac{a^2 x_{1}^{2}}{ r^2}\right)\;,\\ \notag
            \langle T\indices{^{\bar{\phi}}_{\bar{t}}}\rangle_0&=-\frac{3\ell a}{16\pi G_3 r^5}\left(\mu-\frac{q^2\ell}{r}\right)\;.
\end{align}
In the neutral limit we recover the quantum stress-tensor components of the quantum Kerr$_{3}$ black hole \cite{Panella:2024sor}.

\subsubsection*{Thermodynamics}

Let us now briefly turn to the horizon thermodynamics. Since the acceleration horizon is pushed to the boundary, the Euclidean spacetime is regular, upon removing the conical singularity at the location of the horizon in Lorentzian signature. Following the same procedure as in Section \ref{sec:qBHthermo}, we can compute the on-shell Euclidean action, again leading to (\ref{eq:onshellactgeom}), with $\lambda=0$. From the on-shell action, it is straightforward to derive the thermodynamics in the grand canonical ensemble, where we find the mass (\ref{eq:massflat}) identified above as the internal energy, and similarly for the angular momentum, angular velocity, electric charge and conjugate potential. Unsurprisingly, these quantities all obey a semi-classical first law of thermodynamics, where the  generalized entropy has replaced the classical area-entropy variation.

\section{Discussion and outlook}\label{sec:disc}

Our primary focus in this article was to derive the thermodynamics of charged and rotating quantum BTZ black holes. We accomplished this by way of braneworld holography, where the three-dimensional quantum black hole -- an exact solution to an induced theory of semi-classical gravity on the brane -- is totally described by a classical accelerating AdS$_{4}$ black hole that localizes on an end-of-the-world brane. We evaluated the grand canonical partition function by computing the on-shell Euclidean action of the bulk black hole, which subsequently led to the thermodynamic variables of the quantum BTZ black hole. This provides a first principles derivation for the whole qBTZ family, justifying the thermodynamic interpretation of the mass and angular momentum, and serves as a derivation of the generalized entropy of the quantum black hole. 

We then proceeded to construct charged and rotating quantum black holes in dS$_{3}$ and Mink$_{3}$ using ETW Randall-Sundrum branes. Our work is consistent with \cite{Emparan:2022ijy}, showing quantum backreaction effects induce black hole horizons where there were none previously. We subsequently derived the horizon thermodynamics for each black hole, all of which arise from appropriate thermodynamic derivatives of the on-shell Euclidean action of the quantum BTZ black hole in particular limits.  

\vspace{2mm}

\noindent There are a number of avenues worth exploring, some of which we briefly highlight below.

\vspace{2mm}

\noindent \textbf{Breakdown of thermal description.} Here we focused on non-extremal quantum black holes. With charge and rotation, the black holes studied in this work have (near-) extremal limits, where, as the temperature approaches zero, the thermal entropy remains large. This signals a breakdown in the semi-classical thermal description \cite{Preskill:1991tb}. Indeed, for classical non-supersymmetric black holes near-extremality, 1-loop quantum effects play a pivotal role at low-temperatures $T$, yielding $\log(T)$ modifications to the entropy and a gapless microscopic density of states \cite{Iliesiu:2020qvm,Iliesiu:2022onk}, suggesting black holes near-extremality lack a sensible semi-classical geometric description. 

A similar breakdown of the semi-classical description of quantum black holes is expected. Near-extremality, the entropy of the quantum black holes schematically goes like
\beq S_{\text{gen}}^{(3)}=S_{0}^{(3)}+\frac{4\pi^{2}}{M_{\text{gap}}}T+...\;,\eeq
for extremal (generalized) entropy $S_{0}$ as measured in the semi-classical approximation and energy scale $M_{\text{gap}}$ where quantum fluctuations of the geometry become important; the semi-classical
approximation breaks down when $T<M_{\text{gap}}$. The mass gap for static charged quantum BTZ black holes was computed in \cite{Climent:2024nuj}. With rotation, due to an interplay of rotation and quantum backreaction effects, there are two branches of (near-) extremal quantum black holes, leading to two different possible mass gaps $M_{\text{gap}}^{I}$ and $M_{\text{gap}}^{II}$ \cite{Bhattacharya:2025tdn}. Notably, one of the gaps, $M_{\text{gap}}^{I}$, has relatively exotic features, where it is non-monotonic in angular momentum and is not always strictly positive (unlike $M_{\text{gap}}^{II}$). With charge and rotation, there exists a limit where both mass gaps equal zero, suggesting a regime distinct from standard semi-classical black hole thermodynamics such that extremal charged rotating quantum black holes provide a window into deeper quantum gravitational phases. Our evaluation of the on-shell Euclidean action is the first step in performing the necessary perturbative 1-loop corrections to the tree-level thermodynamics derived here, and probe these novel phases, for the qBTZ family and the quantum dS$_{3}$ and Mink$_{3}$ black holes.

\vspace{2mm}

\noindent \textbf{Stability of quantum black holes.} Quantum backreaction effects enrich the thermal phase behavior of the qBTZ family, relative to the classical BTZ black hole. Indeed, the neutral static qBTZ features reentrant phase transitions between (quantum) thermal AdS$_{3}$ and thermally stable quantum black holes \cite{Frassino:2023wpc}.\footnote{See also \cite{Johnson:2023dtf} for an analysis of the specific heat of neutral quantum BTZ black holes.} Including charge and rotation can eliminate the static neutral reentrant phases \cite{Bhattacharya:2025tdn}, however, there is a rich phase structure, especially upon accounting for variable pressure and central charge as introduced in \cite{Frassino:2022zaz}. It would be interesting to explore the thermal phases and stability of the non-AdS quantum black holes constructed here.  Further, while the quantum BTZ family admits thermally stable black holes, they can nonetheless feature (induced) superradiant instabilities, as shown in the neutral rotating case \cite{Cartwright:2025fay}. This stands in stark contrast with the classical BTZ black hole, which, under particular conditions, is linearly stable against perturbations due to a probe field. Building off \cite{Cartwright:2024iwc,Cartwright:2025fay}, it would be worth probing the charged and rotating qBTZ family, and, in particular, the dS and flat quantum black holes. This would develop our understanding of quantum induced superradiance, where, classically a geometry may be linearly and modally stable, yet quantum backreaction alters the geometry to induce dynamical instabilities.

\vspace{2mm}

\noindent \textbf{Cosmic censorship and quantum black holes.} All of the quantum black holes under consideration have curvature singularities, shrouded by an event horizon. This implies that a notion of the weak cosmic censorship conjecture \cite{Penrose:1968ar,Penrose:1969pc} extends to semi-classical gravity. Indeed, it seems quantum backreaction enforces a notion of weak cosmic censorship \cite{Frassino:2025buh}. This is in part evidenced by the fact that the rotating quantum BTZ black hole cannot be overspun \cite{Frassino:2024fin} by sending in a test particle with angular momentum. It would be interesting to further test if charged and rotating quantum black holes can be overcharged or overspun. Preliminary evidence analyzing the sign of the aforementioned mass gap indicates that, upon imposing certain physical requirements, quantum BTZ black holes cannot be overcharged or overspun. A full treatment of this problem is currently underway.

Further, all known quantum black holes in AdS$_{3}$ obey a proposed semi-classical analog of the Penrose inequality \cite{Frassino:2024bjg,Bhattacharya:2025tdn}. The Penrose inequality \cite{Penrose:1973um}, which also lacks a rigorous proof, is an inequality relating the mass of the black hole to the classical horizon area. Since a heuristic derivation of the inequality assumes weak cosmic censorship, it is often thought any violation of the Penrose inequality indicates a violation of cosmic censorship, i.e., the existence of naked singularities. In particular, all charged or rotating quantum BTZ black holes satisfy \cite{Frassino:2024bjg} 
\beq 8\mathcal{G}_{3}(M-M_{\text{cas}})\geq \frac{1}{\ell_{3}^{2}}\left(\frac{2\mathcal{G}_{3}S_{\text{gen}}}{\pi}\right)^{2}\;. \label{eq:QPI}\eeq
On the left-hand side $M_{\text{cas}}=-1/8\mathcal{G}_{3}$ denotes the Casimir contribution to the energy, while on the right-hand side, inspired by the higher-dimensional quantum Penrose inequality \cite{Bousso:2019bkg,Bousso:2019var}, the classical horizon area has been replaced for the generalized entropy. Quantum BTZ black holes with both charge and rotation were also found to obey the conjectured inequality (\ref{eq:QPI}) \cite{Bhattacharya:2025tdn}. Though this observation does not prove a quantum Penrose inequality
exists in (2+1)-dimensional semi-classical gravity, the fact there are no violations (even when backreaction is large and non-perturbative) provides strong circumstantial evidence in its favor.

\vspace{5mm}

\noindent\textbf{Acknowledgments}

\vspace{2mm}

\noindent We are grateful to Robie Hennigar, Emanuele Panella, and Juan Pedraza for thoughtful discussions. AS is supported by the STFC consolidated grant ST/X000753/1 and is partially funded by the Royal Society via the grant ``Concrete Calculables in Quantum de Sitter''.

\appendix

\section{Semi-classical and holographically induced gravity} \label{app:scgravrev}

Here we briefly review the regimes of validity of semi-classical gravity and how it is best understood as an asymptotic expansion. We then provide additional context for induced braneworld gravity. For a longer review on braneworld gravity, see \cite{Bueno:2022log,Panella:2024sor}.

\subsection*{Validity of semi-classical gravity}

Semi-classical gravity is the framework for describing quantum fields in a dynamical curved background. Turning off gravity results in quantum fields in (fixed) curved backgrounds, the original setting for Hawking's derivation of black hole radiation. In addition to the equations of motion for the quantum fields, semi-classical (Einstein) gravity is governed by the semi-classical Einstein equations (\ref{eq:semieineq}).
As it stands, the semi-classical field equations (\ref{eq:semieineq}) are incomplete without further context, as we now briefly describe.

The semi-classical field equations describe how quantum matter backreacts on the classical spacetime metric $g_{ab}$. The matter stress-tensor $T^{\text{mat}}_{ab}$ is  typically quadratic in quantum field operators and is thus ill-defined on account of  being a product of distribution-valued operators. To make sense of the right-hand side of (\ref{eq:semieineq}), one must regulate and renormalize the quantum stress-tensor; this is the only way one can sensibly take the expectation value for an appropriate quantum state that is compatible with describing the spacetime with metric $g_{ab}$ that is free of divergences. By `appropriate quantum state' one technically means a positive linear functional in the sense of algebraic quantum field theory (cf. \cite{Haag:1992hx,Araki:1999ar}). 

Solving the semi-classical Einstein equations is technically much more challenging than solving the classical Einstein equations. Indeed, the first task is to compute the renormalized stress-tensor in a given background. A set of ``axioms'' for a successful renormalization procedure was given by Wald \cite{Wald:1977up,Wald:1978pj}. Various schemes include adiabatic regularization \cite{Fulling:1974zr,Fulling:1974pu,Parker:1974qw} (see also \cite{Birrell:1982ix,Fulling:1989nb}) and Hadamard point-splitting \cite{Brunetti:1999jn,Hollands:2001nf}. Either way, the regularization procedure formally results in an averaged quantum stress-tensor $\langle T^{\text{mat}}_{ab}\rangle$ that depends on covariantly conserved tensors quadratic in the Riemann curvature of the background $g_{ab}$. Logarithmically divergent contributions to the stress-tensor may be absorbed by adding appropriate counterterms to the bare Einstein-Hilbert action quadratic in the curvature. Schematically for $(3+1)$-dimensions,
\beq I_{\text{grav}}=\frac{1}{16\pi G_{N}^{(0)}}\int d^{4}x\sqrt{-g}\left[R-2\Lambda_{0}+\alpha_{0}R^{2}+\beta_{0}R^{2}_{ab}\right]\;.\label{eq:actgravct}\eeq
Appending this gravitational action with the quantum matter action and varying with respect to the metric results in the modified (semi-classical) Einstein's equations, 
\beq G_{ab}+\Lambda_{0} g_{ab}+\alpha_{0}H_{ab}^{(1)}+\beta_{0}H^{(2)}_{ab}=8\pi G^{(0)}_{N}\langle T_{ab}^{\text{mat}}\rangle\;,\label{eq:semieineqapp}\eeq
where $H_{ab}^{(1,2)}$ are  covariantly conserved tensors constructed out of terms quadratic in Ricci curvature and their derivatives.\footnote{Explicitly, 
$$H^{(1)}_{ab}=2\nabla_{a}\nabla_{b}R-2g_{ab}\Box R-\frac{1}{2}g_{qb}R^{2}+2RR_{ab}\;,$$
$$H^{(2)}_{ab}=2\nabla_{c}\nabla_{a}R^{c}_{\;b}-\Box R_{ab}-\frac{1}{2}g_{ab}\Box R-\frac{1}{2}g_{ab}R^{2}_{cd}+2R^{c}_{\;a}R_{cb}\;.$$}
Divergences of $\langle T^{\text{mat}}_{ab}\rangle$ are then removed via an appropriate redefinition of the bare couplings $G_{N}^{(0)},\Lambda_{0},\alpha_{0},\beta_{0}$, the renormalized values of which are then those that are experimentally measured. Subsequently, the renormalized stress-tensor is obtained by subtracting its divergent terms in, e.g., a coincidence limit via point-splitting. It is always possible, however,  to perform additional finite renormalizations of the same form such that the renormalized stress-tensor is uniquely defined up to multiples of the covariantly conserved tensors on the left-hand side of (\ref{eq:semieineqapp}) \cite{Wald:1977up}. 

Solving the semi-classical Einstein equations (\ref{eq:semieineqapp}), therefore, requires one solve a complicated set of non-linear, partial differential equations that are potentially of fourth-order in derivatives of the metric. It is unclear whether such a system will have a well-posed initial value formulation; a unique solution would thus require specifying
the metric and its first three derivatives on a spacelike hypersurface. Moreover, the higher-derivatives can lead to solutions that are unstable with respect to linear metric perturbations, see, e.g., \cite{Horowitz:1978fq,Horowitz:1980fj,Jordan:1987wd}. Precluding the existence of such solutions can be used as a means of restricting the regime of validity of semi-classical gravity \cite{Simon:1990ic,Simon:1990jn}, e.g., the only allowed solutions follow from truncating perturbative expansions at the Planck length squared (see also \cite{Flanagan:1996gw,Anderson:2002fk}). Consequently, the higher-derivative contributions are absorbed into the stress-tensor, leaving only the Einstein tensor, as presented in the main text.

Below we briefly summarize other regimes of validity, settling on a justifiable characterization of semi-classical gravity suitable for our purposes.

.

\vspace{2mm}

\noindent \textbf{Weakly perturbative semi-classical gravity.} In its most naive and narrow view, semi-classical gravity is defined as a perturbative expansion in $G_{N}\hbar$ as $G_{N}\hbar\to0$. This regime is, essentially, QFT in curved space plus weak semi-classical corrections due to the metric deformation coming from the matter sector. 
In this weakly perturbative regime one can, in principle, attempt to solve the semi-classical Einstein equations (\ref{eq:semieineq}) iteratively in powers of $G_{N}\hbar$.
Indeed, backreaction modifications to $(2+1)$-dimensional black holes can be explicitly computed. For example, for a conformally coupled scalar field in an (A)dS$_{3}$ background obeying transparent boundary conditions and in the Hartle-Hawking vacuum, the renormalized stress-tensor (computed using point-splitting regularization and the method of images) has the form \cite{Steif:1993zv,Shiraishi:1993qnr,Lifschytz:1993eb,Martinez:1996uv,Casals:2016ioo,Casals:2019jfo,Emparan:2022ijy,Panella:2023lsi}
\beq \langle T^{a}_{\;b}\rangle=\frac{\hbar F(M)}{8\pi r^{3}}\text{diag}(1,1,-2)\;,\label{eq:stresstensorconfcoupscalar}\eeq
where $F(M)$ is a positive function of the mass.\footnote{The explicit expression for $F(M)$ depends on the background. The radial dependence and diagonal tensorial structure will change for non-transparent boundary conditions \cite{Lifschytz:1993eb}.} Letting this renormalized stress-tensor source the semi-classical Einstein equations results in a (static) geometry
\beq ds^{2}=-f(r)dt^{2}+f^{-1}(r)dr^{2}+r^{2}d\phi^{2}\;,\quad 
 f(r)=\frac{r^2}{L_3^2} - 8G_3M - \frac{2G_{3}\hbar F(M)}{r}\;.
\label{eq:pertbackAdS3geom}\eeq
Notice that the semi-classical correction enters at the order of the Planck scale $L_{\text{P}}^{(3)}=G_{3}\hbar$. Of course, this is the scale when quantum gravitational effects are important and the weakly perturbative semi-classical analysis cannot be trusted. 

An important lesson from this exercise is it exemplifies a way in which semi-classical gravity can be made self-consistent. Should a large $c\gg1$ number of conformally coupled scalars be considered, their combined quantum effect are expected to result in a stress-tensor $\langle T^{\text{mat}}_{ab}\rangle\sim c\hbar$ and a metric correction $\sim Gc\hbar=cL_{\text{P}}\gg L_{\text{P}}$. At the very least, at this scale quantum gravity effects may be safely neglected. Still, the theory is expected to break down when the fluctuations of the stress-tensor become large \cite{Ford:1982wu}. As we now describe, however, ignoring quantum gravitational corrections, even when backreaction effects are strong, can be consistently realized in an appropriate asymptotic expansion.

\vspace{2mm}

\noindent \textbf{Semi-classical gravity as an asymptotic expansion.} Suppose gravity with coupling $G_{N}$ interacts with a large-$c$ amount of light quantum matter fields, which are not necessarily interacting. In a large-$c$ asymptotic expansion with $G_{N}c\hbar$ held fixed, metric fluctuations are subdominant to the matter loops in semi-classical perturbation theory, and the gravitational field may be effectively treated as classical \cite{Hartle:1981zt}. In particular, classical dynamical gravity is preserved in the semi-classical limit:
\beq G_{N}\hbar\to0\;,\quad  \text{with}\quad cG_{N}\hbar\equiv\ell^{d-2} \quad  \text{fixed}\;,\label{eq:scasasymp}\eeq
for length $\ell$ controlling the strength of backreaction and spacetime dimension $d$.

It is only with the asymptotic expansion (\ref{eq:scasasymp}) that the semi-classical Einstein equations (\ref{eq:semieineq}) have, arguably, been derived. The simplest example involves a large-$c$ number of free scalar matter fields conformally coupled to (3+1)-dimensional Einstein gravity \cite{Hartle:1981zt}. 
Further, all known consistent constructions of quantum black holes invoke (\ref{eq:scasasymp}). In particular,  while models of $(1+1)$-dimensional dilaton-gravity cannot be consistently solved in a small-$\hbar$ expansion, the semi-classical field equations are exactly solvable for large-$c$ conformal matter with $cG_{N}\hbar$ fixed, and evaporating black holes can be explicitly constructed \cite{Christensen:1977jc,Callan:1992rs,Russo:1992ax,Bose:1995pz,Fabbri:1995bz,Fabbri:2005mw}. 
This is also the setting for holographic braneworlds and the quantum black holes studied here.

\subsection*{Induced braneworld gravity} Consider a bulk asymptotically $\text{AdS}_{d+1}$ spacetime $\mathcal{M}$ with cosmological constant $\Lambda_{d+1}=-d(d-1)/2\ell_{d+1}^{2}$ governed by classical Einstein-Hilbert gravity  
\beq
I_{\text{bulk}}=\frac{1}{16\pi G_{d+1}}\int_{\mathcal{M}} d^{d+1}x\sqrt{-\hat{g}}\left(\hat{R}-2\Lambda_{d+1}\right)+\frac{1}{8\pi G_{d+1}}\int_{\partial\mathcal{M}}d^{d}x\sqrt{-h}K\;.
\label{eq:BulkTheory}\eeq
Here $G_{d+1}$ is the $d+1$-dimensional Newton's constant, $\hat{g}_{ab}$ is the metric endowed on $\mathcal{M}$. The Gibbons-Hawking-York boundary term is present such that the variational problem is well-posed for boundary submanifold $\partial\mathcal{M}$. Working in the large-$c$, planar-diagram limit, the bulk gravity theory is taken to have a dual holographic description in terms of a $\text{CFT}_{d}$ living on the asymptotic conformal boundary $\partial\mathcal{M}$. 

On-shell, the bulk action will have an infrared divergence as one approaches the asymptotic boundary of AdS. Such IR divergences correspond to UV divergences via the standard holographic dictionary, and can be regulated according to holographic renormalization. In this prescription, the bulk action decomposes into two terms, one which is finite as one approaches the conformal boundary, denoted $I_{\text{fin}}$, and another which diverges, denoted as $I_{\text{div}}$. The divergent contribution can be covariantly cast in terms of curvature invariants of the induced geometry on a codimension-1 IR regulator hypersurface.

In braneworld holography \cite{deHaro:2000wj}, the bulk IR cutoff surface $\partial\mathcal{M}$ is replaced by a $d$-dimensional end-of-the-world brane $\mathcal{B}$ at a small distance away from the AdS$_{d+1}$ boundary. Notably, the brane metric is dynamical, governed by a holographically induced higher-curvature theory of gravity coupled to the $d$-dimensional CFT (now with a cutoff due to the presence of the ETW brane). Precisely, the induced brane theory is found by adding the bulk theory (\ref{eq:BulkTheory}) to the brane action (\ref{eq:genbraneact}), and then integrating out the bulk up to the ETW brane $\mathcal{B}$, as in holographic regularization. The resulting induced theory has the action
\beq
I\equiv I_{\text{Bgrav}}[\mathcal{B}]+I_{\text{CFT}}[\mathcal{B}]\,.
\label{eq:inductheorygen}\eeq
Schematically, the brane gravity theory is (see, e.g., \cite{Chen:2020uac,Bueno:2022log})
\beq 
\begin{split} 
I_{\text{Bgrav}}&=2I_{\text{div}}+I_{\text{brane}}\;,
\end{split}
\label{eq:Ibgravgen}\eeq
where the factor of two accounts for integrating out the bulk on both sides of the brane. 
In general, this theory has an infinite tower of higher curvature densities, entering at order $\ell^{2}_{d+1}$, and has induced Newton and cosmological constants, $G_{d}$ and $\Lambda_{d}$, respectively, 
\beq 
\begin{split} 
&G_{d}=\frac{d-2}{2\ell_{d+1}}G_{d+1}\;,\\
&\Lambda_{d}=\frac{-(d-1)(d-2)}{2L_{d}^{2}}\;,\quad \text{with}\quad \frac{1}{L_{d}^2}=\frac{2}{\ell_{d+1}^2}\left(1-\frac{4\pi G_{d+1}\ell_{d+1}}{d-1}\tau\right)\;.
\label{eq:effGd}
\end{split}
\eeq
Meanwhile, the CFT action $I_{\text{CFT}}[\mathcal{B}]$ characterizing the $d$-dimensional cutoff CFT living on the brane, is identified with the finite contribution to the regulated bulk action upon integrating out the bulk (see, e.g., the discussion around Eq. (3.18) in \cite{Panella:2024sor}). 

In this article, we are interested in scenarios where the bulk is governed by Einstein-Maxwell-AdS$_{4}$ gravity. In this case, the induced theory on the brane is \cite{Climent:2024nuj}
\beq I=\frac{1}{16\pi G_{3}}\int_{\mathcal{B}}d^{3}x\sqrt{-h}\left[R-2\Lambda_{3}+\ell_{4}^{2}\left(\frac{3}{8}R^{2}-R_{ij}^{2}\right)+...\right]+I_{\text{EM}}+I_{\text{CFT}}\;,\label{eq:inducactqbtzchar}\eeq
where the electromagnetic term $I_{\text{EM}}$ is
equal to the local counterterms associated with the four-dimensional bulk Maxwell action that are included in holographic renormalization \cite{Taylor:2000xw}
\beq
\begin{split}
 I^{\text{ct}}_{\text{EM}}=\frac{\ell_{4}\ell^{2}_{\ast}}{8\pi G_{4}}&\int d^{3}x\sqrt{-h}\biggr[-\frac{5}{16}F_{jk}^{2}+\ell_{4}^{2}\biggr(\frac{1}{288}RF_{jk}^{2}-\frac{5}{8}R^{i}_{\;j}F_{ik}F^{jk}\\
&+\frac{3}{98}F^{ij}(\nabla_{j}\nabla^{k}F_{ki}-\nabla_{i}\nabla^{k}F_{kj})+\frac{5}{24}\nabla_{i}F^{ij}\nabla_{k}F^{k}_{\;j}\biggr)+\mathcal{O}(L_{4}^{3})\biggr]\;.
\end{split}
\eeq
Treating $\ell<\ell_{3}$ such that $\ell_{4}\sim \ell$, the effective theory on the brane becomes
\beq
\begin{split}
 I&=\frac{1}{16\pi G_{3}}\int d^{3}x\sqrt{-h}\biggr[R+\frac{2}{\ell_{3}^{2}}-\frac{\tilde{\ell}_{\ast}^{2}}{4}F_{jk}^{2}\\
&+\ell^{2}\left(\frac{3}{8}R^{2}-R_{ij}^{2}\right)+\frac{4}{5}\ell^{2}\tilde{\ell}^{2}_{\ast}\biggr(\frac{1}{288}RF^{2}-\frac{5}{8}R^{i}_{\;j}F_{ik}F^{jk}\\
&+\frac{3}{98}F^{ij}(\nabla_{j}\nabla^{k}F_{ki}-\nabla_{i}\nabla^{k}F_{kj})+\frac{5}{24}\nabla_{i}F^{ij}\nabla_{k}F^{k}_{\;j}\biggr)+\mathcal{O}(\ell^{3})\biggr]+I_{\text{CFT}}\;.
\end{split}
\label{eq:indactbranecharge}\eeq
Recall $\ell\sim c_{3}G_{3}\hbar$ for (large) CFT central charge $c_{3}$, and induced Newton constant $G_{3}$. We see then that braneworld holography naturally organizes the induced semi-classical theory as an asymptotic expansion in $c_{3}G_{3}\hbar$, consistent with (\ref{eq:scasasymp}). Comparing to the action (\ref{eq:actgravct}), we see that the higher-derivative contributions here serve as appropriate counterterms to regularize the averaged quantum stress-tensor, for a specific choice of $\alpha_{0}$ and $\beta_{0}$.

\section{C-metric geometry} \label{app:cmetgeom}

Here we list another coordinate system for the AdS$_{4}$ C-metric that proves convenient for practical computations involving the on-shell action. 

\vspace{2mm}

\noindent \textbf{Lorentzian signature.} Take the charged and rotating C-metric (\ref{eq:rotatingCmetqbtz}) and perform the following coordinate transformation and parameter identifications
\beq
\begin{split} 
&t\to t\ell\;,\quad r\to -\frac{\ell}{y}\;,\\
&\lambda\equiv \frac{\ell^{2}}{\ell_{4}^{2}}-1=\frac{\ell^{2}}{\ell_{3}^{2}}\;,\quad k=-\kappa\;, \quad a\to a\ell \;.
\end{split}
\label{eq:xytransapp}\eeq
Here $t$, $y$, and $a$ are all dimensionless.  The resulting geometry is
\beq 
\begin{split}
ds^{2}&=\frac{\ell^2}{(x-y)^{2}}\biggr[-\frac{H(y)}{\Sigma(x,y)}(dt+ax^{2}d\phi)^{2}+\frac{\Sigma(x,y)}{H(y)}dy^{2}+\frac{\Sigma(x,y)}{G(x)}dx^{2}+\frac{G(x)}{\Sigma(x,y)}(d\phi-ay^{2}dt)^{2}\biggr]\;,
\end{split}
\label{eq:rotCmetapp}\eeq
with metric functions
\beq
\begin{split}
&H(y)=\lambda-ky^{2}+\mu y^{3}+a^{2}y^{4}+q^{2}y^{4}\;,\quad \Sigma(x,y)=1+a^{2}x^{2}y^{2}\\
&G(x)=1+kx^{2}-\mu x^{3}+a^{2}\lambda x^{4}-q^{2}x^{4}\;.
\end{split}
\eeq
The gauge field, meanwhile, becomes
\beq A=A_{a}dx^{a}=\frac{2\ell q y}{\ell_{\star}(1+a^{2}x^{2}y^{2})}(dt+ax^{2}d\phi)\;.\eeq

 We fix the period of the angular variable $\phi$ such that $\phi \sim \phi+\Delta\phi$ with $\Delta\phi=4\pi/|G'(x_{1})|$ to avoid a conical singularity at $x=x_{1}$.
The $x,y$ coordinate ranges are restricted to be $0\leq x\leq x_{1}$, and $-\infty\leq y\leq x$, where  $y=-\infty$ corresponds to a curvature singularity hidden behind the bulk horizon, which we denote as $y=y_{+}$.  The region $x,y\to0$ corresponds to an asymptotic region far from the black hole. 

Written in these coordinates, it is clear the metric (\ref{eq:rotCmetapp}) has umbilic surfaces at $x=0$ and $y=0$, for which 
\beq K_{ij}^{(x)}=-\frac{1}{\ell}h_{ij}^{(x)}\;,\quad K_{ij}^{(y)}=-\frac{\sqrt{\lambda}}{\ell}h^{(y)}_{ij}\;,\eeq 
where $h^{(x)}_{ij}$ and $K^{(x)}_{ij}$ respectively denote the induced metric and extrinsic curvature at the $x=0$ hypersurface; similarly for $h_{ij}^{(y)}$ and $K_{ij}^{(y)}$.\footnote{Note that the spacelike unit normals for constant-$x$ and constant-$y$ hypersurfaces are, respectively, $n^{(x)}_{a}=\ell|x-y|^{-1}\sqrt{\frac{\Sigma}{G}}\delta^{x}_{a}$ and $n^{(y)}_{a}=\ell|x-y|^{-1}\sqrt{\frac{\Sigma}{H}}\delta^{y}_{a}$.} 
Substituting this into Israel's junction conditions (\ref{eq:israeljuncconds}), assuming both brane actions have the purely tensional form (\ref{eq:genbraneact}) gives the tensions (\ref{eq:branetensionsxy}).

To impose bulk regularity, one must move to the barred coordinates $(\bar{t},\bar{\phi})$ in (\ref{eq:transKillvec}) where under (\ref{eq:xytransapp}) we have
\beq t\to t\ell=\eta\ell(\bar{t}-\bar{a}\ell_{3}\bar{\phi})\;,\quad \phi\to \eta \left(\bar{\phi}-\frac{\bar{a}\bar{t}\ell^{2}}{\ell_{3}}\right)\;,\eeq
where $\bar{t}$ is dimensionless and $\bar{a}$ has dimensions of inverse length. Inverting, we find
\beq \bar{t}=\frac{1}{(1-\bar{a}^{2}\ell^{2})\eta}(t+\bar{a}\ell_{3}\phi)\;,\quad \bar{\phi}= \frac{1}{(1-\bar{a}^{2}\ell^{2})\eta}\left(\phi+\frac{\bar{a}\ell^{2}}{\ell_{3}}t\right)\;.\label{eq:barcoordapp}\eeq
It is easy to verify that since $t\sim t -2\pi \tilde{a}\ell_{3}\eta$ and $\phi\sim \phi+2\pi \eta$ (along orbits of the rotational Killing vector),  it follows $\bar{t}\sim \bar{t}$ and $\bar{\phi}\sim \bar{\phi}+2\pi$. We can further introduce $\bar{y}$ via 
\beq \frac{\ell^{2}}{\bar{y}^{2}}=\frac{\ell^{2}}{y^{2}}(1-\bar{a}^{2}\ell^{2})\eta^{2}+\frac{\ell^{2}}{y_{s}^{2}}\;,\quad y_{s}^{2}=\left(\frac{\bar{a}^{2}\ell_{3}^{2}\eta^{2}}{x_{1}^{2}}(2+kx_{1}^{2})\right)^{-1}\;.\eeq

\vspace{2mm}

\noindent \textbf{Euclidean signature.} Via the  Wick rotations $t\to -it_{E}$, $a\to ia_{E}$, and $q\to iq_{E}$ 
for Euclidean time $t_{E}$ and $a_{E},q_{E}\in\mathbb{R}$, the metric (\ref{eq:rotCmetapp}) becomes (\ref{eq:rotCmetEuc}), which we report here again for convenience,
\beq 
\begin{split}
ds^{2}_{E}&=\frac{\ell^2}{(x-y)^{2}}\biggr[\frac{H(y)}{\Sigma(x,y)}(dt_{E}-a_{E}x^{2}d\phi)^{2}+\frac{\Sigma(x,y)}{H(y)}dy^{2}+\frac{\Sigma(x,y)}{G(x)}dx^{2}+\frac{G(x)}{\Sigma(x,y)}(d\phi-a_{E}y^{2}dt_{E})^{2}\biggr]\;,
\end{split}
\label{eq:rotCmetEucapp}\eeq
with metric functions
\beq
\begin{split}
&H(y)=\lambda-ky^{2}+\mu y^{3}-a_{E}^{2}y^{4}-q_{E}^{2}y^{4}\;,\quad \Sigma(x,y)=1-a_{E}^{2}x^{2}y^{2}\\
&G(x)=1+kx^{2}-\mu x^{3}-a_{E}^{2}\lambda x^{4}+q_{E}^{2}x^{4}\;.
\end{split}
\eeq
The gauge field is
\beq A_{E}=A_{a}dx^{a}=\frac{2\ell q_{E} y}{\ell_{\star}(1-a_{E}^{2}x^{2}y^{2})}(dt_{E}-a_{E}x^{2}d\phi)\;.\eeq 
Note that the norm of the gauge field 
 \beq A_{a}A^{a}=\frac{4q_{E}^{2}}{\ell_{\star}^{2}}\frac{(x-y)^{2}y^{2}\Sigma(x,y)}{(1-a_{E}^{2}x^{2}y^{2})^{2}H(y)}\;,\eeq
 diverges at the horizon. The divergence can be removed by adding to the field a term $A_{E}\to A_{E}+C_{E}dt_{E}$ with $C_{E}\equiv -2q_{E}y_{+}\ell/\ell_{\star}$. 

 The barred coordinates (\ref{eq:barcoordapp}) rotate to
\beq 
\begin{split} 
&\bar{t}\to -i\bar{t}_{E}\;,\quad \bar{t}_{E}\equiv \frac{1}{(1+\bar{a}_{E}^{2}\ell^{2})\eta}t_{E}-\frac{\bar{a}_{E} \ell_{3}}{(1+\bar{a}_{E}^{2}\ell^{2})\eta}\phi\;,\\
&\bar{\phi}\to \frac{\bar{a}_{E}\ell^{2}}{\ell_{3}(1+\bar{a}_{E}^{2}\ell^{2})\eta}t_{E}+\frac{1}{(1+\bar{a}_{E}^{2}\ell^{2})\eta}\phi\;.
\end{split}
\label{eq:barrcoordEucapp}\eeq
Inverting, we have 
\beq t_{E}=\eta (\bar{t}_{E}+\bar{a}_{E}\ell_{3}\bar{\phi})\;,\quad \phi=\eta\left(\bar{\phi}-\frac{\bar{a}_{E}\ell^{2}}{\ell_{3}}\bar{t}_{E}\right)\;,\label{eq:tephiinbar}\eeq
where we note the gauge potential has the form
\beq A_{E}=\frac{2\ell q_{E}y\eta}{\ell_{\star}(1-a^{2}_{E}x^{2}y^{2})}\left[d\bar{t}_{E}\left(1+\frac{a_{E}x^{2}\bar{a}_{E}\ell^{2}}{\ell_{3}}\right)+d\bar{\phi}\left(\bar{a}_{E}\ell_{3}-a_{E}x^{2}\right)\right]\;.\label{eq:gaugpotbarrapp}\eeq

To ensure the Riemannian geometry is regular requires two identifications. Firstly, there is the one needed to remove the conical singularities arising from the zero $x=x_{1}$ of $G(x)$, 
such that 
\beq (\bar{t}_{E},\bar{\phi})\sim (\bar{t}_{E},\bar{\phi}+2\pi)\;.\eeq
It is easy to verify this follows from $t_{E}\sim t_{E}+2\pi \bar{a}_{E}\ell_{3}\eta$ and $\phi\sim \phi+2\pi\eta$. Further, the Euclidean geometry has a conical singularity at the horizon $y=y_{+}$, where $H(y_{+})=0$. 

To see this, recall the near-horizon region of the Euclidean black hole has line element (\ref{eq:nearhorqbh}).
Notice the parenthetic term has the form of a two-dimensional disk $C_{2}$ attached to the horizon, with metric 
\beq ds^{2}_{C_{2}}=d\rho^{2}+\frac{(H'(y_{+}))^{2}\rho^{2}}{4\Sigma_{+}^{2}}d\chi^{2}\;,\eeq
where we introduced angular coordinate 
\beq 
\begin{split}
\chi&\equiv t_{E}-a_{E}x^{2}\phi\\
&=\eta \bar{t}_{E}\left(1+\frac{a_{E}x^{2}\bar{a}_{E}\ell^{2}}{\ell_{3}}\right)+\eta\bar{\phi}(\bar{a}_{E}\ell_{3}-a_{E}x^{2})\;.
\end{split}
\eeq
We encounter a conical singularity at the horizon ($\rho=0$) unless we periodically identify 
\beq \chi \sim \chi+\frac{4\pi\Sigma_{+}}{H'(y_{+})}\;.\label{eq:periodrot}\eeq
Equivalently, we require the periodicities $t_{E}\sim t_{E}+\Delta t_{E}$ and $\phi\sim \phi+\delta\phi$. In $(\bar t, \bar \phi)$ coordinates, the periodicities are $\bar t_E \sim \bar t_E+\beta$ and $\bar \phi \sim \bar \phi -i\Omega \beta$. From identification (\ref{eq:periodrot}) we have
\beq 
\begin{split}
\frac{4\pi \Sigma_{+}}{H'(y_{+})}&=\Delta t_{E}-a_{E}x^{2}\delta\phi\\
&=\eta \Delta \bar{t}_{E}\left(1+\frac{a_{E}x^{2}\bar{a}_{E}\ell^{2}}{\ell_{3}}\right)+\eta\delta\bar{\phi}(\bar{a}_{E}\ell_{3}-a_{E}x^{2})\;.
\end{split}
\label{eq:idencone}\eeq
where $\delta\phi\neq \Delta\phi$, and we introduced periodicites $\bar{t}_{E}\sim \bar{t}_{E}+\Delta\bar{t}_{E}$ and $\bar{\phi}\sim \bar{\phi}+\delta\bar{\phi}$.

\vspace{2mm}

\noindent \textbf{Grand canonical data.} From the twisted KMS condition, $(\bar{t}_{E}\ell,\bar{\phi})\sim (\bar{t}_{E}\ell+\beta,\bar{\phi}-i\beta\Omega)$, we identify $\Delta\bar{t}_{E}\ell=\beta$ and $\delta\bar{\phi}=-i\Omega \beta$.\footnote{Ordinarily, inverse temperature has dimensions of length. Here we have $\bar{t}_{E}\ell$ for dimensionless time $\bar{t}_{E}$, hence the appearance of $\ell$.} Together with (\ref{eq:idencone}), we find
\beq \frac{4\pi}{H'(y_{+})}(1-a_{E}^{2}x^{2}y_{+}^{2})=\eta\beta\ell^{-1}\left(1+\frac{a_{E}x^{2}\bar{a}_{E}\ell^{2}}{\ell_{3}}\right)-i\Omega\beta \eta(\bar{a}_{E}\ell_{3}-a_{E}x^{2})\;.\eeq
This relation is valid for any fixed value of $x$. In particular, for $x=x_{1}$ the second term cancels, and we find the inverse temperature is given by (\ref{eq:invtemp}).
Meanwhile, for fixed $x\neq x_{1}$, we compare all terms proportional to $x^{2}$ and identify $\Omega$ as (\ref{eq:Omeucc}). 

From coordinates (\ref{eq:tephiinbar}), combined with $t_{E}\sim t_{E}+\Delta t_{E}$, we are able to read off 
\beq
\begin{split}
\Delta t_{E}&=\frac{\eta\beta}{\ell}(1-i\bar{a}_{E}\ell\ell_{3}\Omega)\\
&=\frac{4\pi}{H'(y_{+})}\;.
\end{split}
\eeq
In terms of geometric quantities, period $\Delta t_{E}$ has the same form as the static, neutral black hole (cf. Eq. (E.6) of \cite{Panella:2024sor}). Incidentally, this implies $\delta \phi=a_{E}y_{+}^{2}\Delta t_{E}$, which we could have arrived at by demanding the angular variable $\psi\equiv \phi-a_{E}y_{+}^{2}t_{E}$ to satisfy $\psi\sim \psi$ under $\phi\sim \phi+\delta\phi$ and $t_{E}\sim t_{E}+\Delta t_{E}$.

\vspace{2mm}

\noindent \textbf{Constrained thermodynamic variables.} We derive the thermodynamic variables for the quantum black holes via  the on-shell Euclidean action. As reported in the main text, we know the on-shell action and potentials $\{\beta,\Omega,\Phi\}$ as functions of the variables $\{z,\alpha,\tilde{q}\}$. Thus, to determine the thermodynamic variables $\{E,J,Q,S\}$ we must evaluate the following constrained derivatives (\ref{eq:thermoquantsgcens})  
\beq 
\begin{split}
&J= -\frac{1}{\beta}\left(\frac{\partial I_{E}^{\text{on-shell}}}{\partial\Omega}\right)_{\hspace{-1mm}\beta,\Phi}\;,\quad Q=-\frac{1}{\beta}\left(\frac{\partial I_{E}^{\text{on-shell}}}{\partial\Phi}\right)_{\hspace{-1mm}\beta,\Omega}\;,\\
&E=\left(\frac{\partial I_{E}^{\text{on-shell}}}{\partial\beta}\right)_{\hspace{-1mm}\Omega,\Phi}+\Omega J+\Phi Q\;,\quad S=\beta \left(\frac{\partial I_{E}^{\text{on-shell}}}{\partial\beta}\right)_{\hspace{-1mm}\Omega,\Phi}-I_{E}^{\text{on-shell}}\;.
\end{split}
\label{eq:thermoquantsgcensapp}\eeq
Each derivative is evaluated using standard formulae involving Jacobians, 
\beq \left(\frac{\partial f}{\partial g}\right)_{h,p}=\frac{\left(\frac{\partial(f,h,p)}{\partial(w^{1},w^{2},w^{3})}\right)}{\left(\frac{\partial(g,h,p)}{\partial(w^{1},w^{2},w^{3})}\right)}\;,\eeq
with 
\beq \frac{\partial(f,h,p)}{\partial(w^{1},w^{2},w^{3})}\equiv \text{det}\begin{pmatrix} \partial_{w^{1}}f&\partial_{w^{2}}f&\partial_{w^{3}}f\\ \partial_{w^{1}}h&\partial_{w^{2}}h&\partial_{w^{3}}h\\ \partial_{w^{1}}p&\partial_{w^{2}}p&\partial_{w^{3}}p\end{pmatrix}\;,\eeq
for real functions $f(w^{1},w^{2},w^{3})$, $g(w^{1},w^{2},w^{3})$ and $p(w^{1},w^{2},w^{3})$.  

For example, 
\beq \left(\frac{\partial I_{E}^{\text{on-shell}}}{\partial \beta}\right)_{\Omega,\Phi}=\frac{\left(\frac{\partial(I_{E}^{\text{on-shell}},\Omega,\Phi)}{\partial(z,\alpha,\tilde{q})}\right)}{\left(\frac{\partial(\beta,\Omega,\Phi)}{\partial(z,\alpha,\tilde{q})}\right)}\;,\eeq
and similarly for the other variables.

\bibliography{qBHrefs}

\end{document}